\documentclass[twocolumn,showpacs,aps,pra,amsmath,amssymb,floatfix]{revtex4}\preprint{draft 1.5}
\usepackage{graphicx}
\usepackage{dcolumn}
\usepackage{bm}
\usepackage{amssymb}
\usepackage{setspace}

\begin{document}

\title{Femtosecond pulses and dynamics of molecular photoexcitation: RbCs example}

\author{B. E. Londo\~no$^{1,2,4}$}
\author{A. Derevianko$^{3}$}
\author{J. E. Mahecha$^{1}$}
\author{A. Crubellier$^{2}$}
\author{E. Luc-Koenig$^{2}$}
\email{Eliane.Luc@lac.u-psud.fr}

\affiliation{$^{1}$Instituto de f\'{\i}sica, Universidad de Antioquia, Calle 67 No 53-108, AA 1226, Medellin, Colombia}
\affiliation{$^{2}$Laboratoire Aim\'e Cotton, CNRS, B\^atiment 505, Universit\'e Paris-Sud 11, 91405  Orsay Cedex, France}
\affiliation{$^{3}$Department of Physics, University of Nevada, Reno, Nevada 89557, USA }
\affiliation{$^{4}$Present address: Facultad de Ciencias, Universidad Antonio Nari\~no, Carrera 3 este No. 47A - 15, Bogota, Colombia }
\date{\today}
\begin{abstract}
We investigate the dynamics of molecular photoexcitation by unchirped femtosecond laser pulses using RbCs as a model system. 
This study is motivated by a goal of optimizing a two-color scheme of transferring vibrationally-excited ultracold molecules to their absolute ground state. In this scheme  
the molecules are initially  produced by photoassociation or magnetoassociation in bound vibrational levels close to the first dissociation threshold. We analyze here the first step of the two-color path as a function of pulse intensity from the low-field to the high-field regime. We use two different approaches, a global one, the 'Wavepacket' method, and a restricted one, the 'Level by Level' method where the number of vibrational levels is limited to a small subset. The comparison between the results of the two approaches allows one to gain qualitative insights into the complex dynamics of the high-field regime. In particular, we emphasize the non-trivial and important role of far-from-resonance levels which are adiabatically excited through  'vertical' transitions with a large Franck-Condon factor. We also point out spectacular excitation blockade due to the presence of a quasi-degenerate level in the lower electronic state. We conclude that selective transfer with femtosecond pulses is  possible in the low-field regime only. Finally, we extend our single-pulse analysis and examine population transfer induced by coherent trains of low-intensity femtosecond pulses. 
\end{abstract}

\pacs{ 33.80.-b, 34.80.Gs, 31.10.+z, 33.15.-e,}

\maketitle
\noindent
\section{\label{sec:intro} Introduction}
\begin{figure}[h]
\centering
\includegraphics[viewport=0 0 613 756, width=8cm, clip]{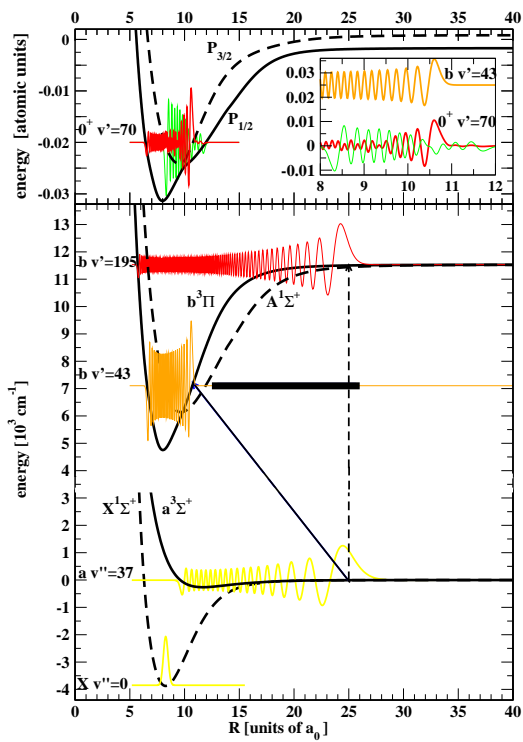}
\caption{\label{ch5:fig:3-levels}(Color online) Photoexcitation of RbCs. Lower panel: Diabatic potentials for the stable $X^1\Sigma^+$ (lower thick dashed black line), the metastable $a^3 \Sigma ^+$ (lower thick continuous black line) and the excited $b^3\Pi$ (upper thick continuous black line) and $A^1\Sigma^+$ (upper thick dashed black line) electronic states. The origin of energy is set at the dissociation limit Rb(5$s$)Cs(6$s$). Vibrational wavefunctions  for the initial level $a^3\Sigma^+$ $v"=37$ (upper continuous yellow [clear-gray]  line) and for the final level $X^1\Sigma^+$ $v"=0$ (lower continuous yellow [clear-gray]  line) in the two-color process. Wavefunctions  for the $b^3\Pi$ $v'=43$  level (continuous orange [dark-gray] line) corresponding to the resonant transition   ('oblique' black  continuous line) and for the $b^3\Pi$ $v'=195$  level (continuous red [black] line) corresponding to the off-resonant `vertical' transition (vertical dashed black line). The wavefunctions are drawn at their absolute energy.  The bandwidth of the laser pulse is shown as the thick black horizontal line. Upper panel: Adiabatic potentials  $0^+ \, P_{1/2}$ (thick continuous black line) and $0^+ \, P_{3/2}$ (thick dashed black line). The energy origin is at the Rb(5$s$)Cs(6$p$) dissociation limit. The two vibrational components of the coupled $0^+ v'=70$ level in the Hund's case $c$ representation are drawn, respectively  $b^3\Pi$  (medium-thick red [black] line)   and $A^1\Sigma^+$ (thin green [medium-gray] line). In the inset,  these two components are compared with the wavefunction of the $b^3\Pi$ $v'=43$ level (medium orange [dark-gray] line) for  $8<R<12$ a$_0$.}
\end{figure}

\noindent
Rb and Cs atoms have been simultaneously trapped and laser cooled in a magneto-optic trap down to  ultracold temperature ($\sim 100 \mu$K). Ultracold RbCs molecules have been formed through photoassociation in excited vibrational levels of the Rb(5$s$)Cs(6$p_{1/2}$) $0^+$, $0^-$ or $1$ symme\-tries. These molecules decay through spontaneous emission, mainly toward stable levels of the Rb(5$s$)Cs(6$s$) $a^3\Sigma^+$ electronic state; the upper of those levels has a binding energy in the range of $5$ cm$^{-1}$  \cite{kerman2004a}. The relevant molecular terms are shown in Fig.~\ref{ch5:fig:3-levels}.

\noindent
In the heteronuclear RbCs molecule, two-step conversion processes from the $a^3\Sigma^+$ state (denoted below by `$a$') toward the $X^1\Sigma^+$ state  (denoted below by `$X$') are possible by using, as intermediate step, levels of the  $0^+$ or $1$ symmetries, with a spin-mixed character. As a result, molecules in the absolute ground level Rb(5$s$)Cs(6$s$) $X^1\Sigma^+$ $v"=0$ are formed. These processes have been recently investigated experimentally \cite{sage2005,kerman2004a} and theoretically \cite{bergeman2004,stwalley2004,tscherneck2007}. 

\noindent
Ultracold stable polar molecules in their absolute ground vibrational level have been populated for the first time \cite{sage2005,kerman2004a} using a two-color incoherent population transfer through a low-lying level of the $1$ state. A resonant `pump' laser pulse transfers the population of the metastable, vibrationally excited $a^3\Sigma^+$ molecules to an electronically excited level; then a second tunable `dump' laser pulse resonantly drives the population to the absolute ground level. The two laser pulses used in this stimulated transfer have a duration of about 5 ns.

\noindent
In the KRb molecule, using a Stimulated Raman Adiabatic Passage (STIRAP) with counterin\-tui\-tive pulses in the microsecond range, Ni {\it et al}~\cite{ni2008} transferred extremely weakly-bound Feshbach molecules in the $a$ electronic state toward the lowest vibrational level either of the stable $X$ or of the metastable $a$ states using intermediate level with symmetry $1$.

\noindent
For several years, researchers at Aim\'e Cotton Laboratory are exploring theoretically, on the example of the Cs$_2$ and Rb$_2$ molecules, coherent schemes using chirped laser pulses to form molecules in an excited electronic state through photoassociation of ultracold atoms, and then to stabilize them through stimulated emission \cite{luc2009, luc2003, luc2004}. The motivation was to fully exploit optical techniques for controlling the formation of cold molecules in the absolute ground level. The studied laser pulses were in the picosecond range, the domain well-adapted to the vibrational dynamics of the wavepackets created by the  pulse in the light-coupled electronic states.  

\noindent
However, from a technological point of view, picosecond lasers and corresponding pulse shapers are not yet available. On the other hand, in the femtosecond domain  there were important recent developments of efficient laser sources and pulse shapers. Furthermore, coherent trains of pulses, obtained from  mode-locked femtosecond lasers \cite{cundiff2002}, permit a transient coherent accumulation of population, manifested by the enhancement of transition probabilities and by a gain in the spectral resolution \cite{stowe2006}.

\noindent
Our objective here is to analyze the possibilities offered by  femtosecond sources in implementing efficient two-color paths for transferring  vibrationally-excited ultracold molecules to their absolute ground state. In this scheme  
the molecules are initially  produced by photoassociation or magnetoassociation in bound vibrational levels close to the first dissociation threshold.
Numerical analysis is carried out for the RbCs molecule.
More precisely, the present paper is devoted to the choice of the optimal pulse for implementing the first step of the two-color paths. Notice that femtosecond pulses have a broad bandwidth and may reach high intensities. Consequently we have to analyze the dynamics of  coherent excitation of  a large number of  vibrational levels, from the low field up to the high field regime. 

\noindent
To solve the time-dependent Schr\"{o}dinger equation, we first use the `Wavepacket' method (WP), where we calculate globally the evolution of  vibrational wavepackets propagating along electronic states coupled by the laser pulse \cite{luc2009}. Using this approach, it appears that, in the high-field regime, the calculated dynamics and the population transfer drastically differ  from what is expected from intuitive two-level-system arguments. To understand these surprising results, we compare the WP results to solutions obtained using a small subset of vibrational levels: we refer to this model as the `Level by Level' method (LbyL). In both approaches, the dependence of the wave function on the interatomic distance $R$ is obtained from the Mapped Fourier Grid Hamiltonian (MFGH) method \cite{kokoouline1999,willner2004}. 

\noindent 
By comparing the WP results with the LbyL solutions, we precisely identify vibrational levels critically responsible for the strongly nonlinear dy\-na\-mics in the high-field regime. 
In the high-field regime, the dynamics of the photoexcitation process is governed  both by nearly-resonant and by far-from-resonance excitations. The adiabaticy  of  the resonant and non-resonant excitations can be easily analyzed in detail in the simple case of a two-level system. For a multilevel system, we show that, in the high-field regime, the dynamics of time-evolution of the population in nearly-resonant levels is strongly affected by the adiabatic excitation of far-from-resonance levels. For a particular level, the adiabaticity  of the excitation by an unshaped Gaussian pulse is  found to be simply related to the value of its detuning with respect to the carrier laser frequency.
In the photoexcitation process under study, the initial level lies close to the dissociation threshold, in an energy domain where the density of vibrational levels is high. We show that the presence of such a quasi-degenerate group of levels in the ground electronic state leads in the high field regime to a spectacular blockade of the excitation process.

\noindent 
We conclude from the  analysis that, while femtosecond laser pulses are concerned, control of the photoexcitation process is possible only in the low field regime. To improve the efficiency of the population transfer, we  investigate some schemes using coherent trains of low-intensity femtosecond pulses. 

\noindent
The paper is organized as follows. First we specify the photoexcitation process (Sec. \ref{ch5:subsec:reaction}) and also  characterize  Gaussian pulse (Sec. \ref{ch5:subsec:pulse}).  Then we briefly describe the two employed approaches (the WP and LbyL methods) to solving the time-dependent Schr\"{o}dinger equation (Sec. \ref{ch5:subsec:TimeD-photEx}).  The photoexcitation dynamics is dramatically affected as the pulse intensity is increased. It's dependence on the pulse intensity is computed  in the WP approach and is described in Sec. \ref{ch5:sec:pi-pulse}. These results are further analyzed in Sec. \ref{ch5:sec:LbyLanalysis} in the framework of the LbyL method. 
This framework allows us to identify levels responsible for the observed photoexcitation dynamics (Sec. \ref{ch5:subsec:levels}). We further exhibit the link between adiabaticity and detuning first in the simple case of a two-level system (Sec. \ref{ch5:subsec:two-level-s}) and then for the multi-level system under study (Sec. \ref{ch5:subsec:LbyL-degenerate}). The excitation blockade due to the presence of quasi-degenerate group of levels in the ground state is studied in Section \ref{ch5:subsubsec:analysis-quasi-degenerate}. Finally, we comment on the photoexcitation dynamics induced by coherent trains of low-intensity femtosecond pulses in Section \ref{ch5:sec:ImproveT}.

\noindent
The paper contains several appendices used for  
recapitulating essential results and to precise notation. 
Appendix \ref{app:MFGH} briefly reviews  the Mapped Fourier Grid Hamiltonian (MFGH) employed throughout the paper. The 'Wavepacket' and the 'Level by Level' methods are described in the Appendix \ref{app:dyn}. Appendix \ref{app:multilevel} recalls the definition of the diabatic and adiabatic bases used in our analysis. A simple model for the blockade of excitation due to the presence of a quasi-degenerate group of levels in the lower electronic state is described in Appendix \ref{ch5:subsec:N-levels}, whereas
Appendix \ref{app:pulse-train} lists relevant properties of ultrashort pulse trains.

\section{\label{ch5:sec:photo-reaction}Photoexcitation of R$\mathbf{b}$C$\mathbf{s}$}
\subsection{\label{ch5:subsec:reaction}Photoexcitation process}
\noindent
In the RbCs molecule, it has been shown that the two-color path $a^3\Sigma^+ v"=37 \rightarrow 0^+ v'=70 \rightarrow X ^1\Sigma^+ v"=0$ is very efficient in transferring to the absolute ground level $X ^1\Sigma^+ v"=0$ the molecules obtained in the $a v"=37$ level after photoassociation followed by spontaneous radiative decay \cite{londono2009}. The $0^+$ symmetry results from the coupling through the spin-orbit interaction of the singlet $A^1\Sigma^+$ and the triplet $b^3\Pi$ electronic states.  The $0^+ v'=70$ level is a mix of vibrational levels $b^3\Pi v'$ (52.7\%) with $v'\sim 43$ and of $A^1\Sigma^+v'$ levels (47.3\%) with $v'\sim25$. In the first step of the two-color path, only the $|b^3\Pi v'\rangle$ components of the coupled wave functions $|0^+ v'\rangle$ can be excited; we have shown that the excitation probabilities $a v"=37 \rightarrow bv'=43$ and $a v"=37 \rightarrow 0^+ v'=70$ level are very similar. Therefore, in this paper, we restrict the analysis of the photoexcitation dynamics to the study of the  $a^3\Sigma^+ v" \rightarrow b^3\Pi v'$ transition. The rotational structure of the vibrational levels as well as the hyperfine structure are ignored.

\noindent
We consider  excitation by a Gaussian laser pulse with a  duration $\tau_L$ and a carrier frequency $\omega_L/2\pi$  resonant with the transition between the vibrational levels $a v"_0=37$ and $ b^3\Pi v_0'=43$, 
\begin{eqnarray}
\hbar\omega_L={E}(b^3\Pi  v_0'=43) - E(a^3\Sigma^+ v"_0=37),
\label{ch5:eq:excit}
\end{eqnarray}
where  $E(a^3\Sigma^+ v"_0=37)$ and ${E}(b^3\Pi  v_0'=43)$ are absolute energies of the two levels.

\noindent
The initial level has a binding energy of only $5.52$ cm$^{-1}$ and it lies very close to the Rb(5$s$)Cs(6$s$) dissociation limit.  
The excited level  with binding energy $4392$ cm$^{-1}$ with respect to the Rb(5$s$)Cs(6$p)$ dissociation limit is tightly bound (Fig.~\ref{ch5:fig:3-levels}).
There are substantial differences in the two vibrational wave functions. 
The wave function of the initial level $ a^3\Sigma^+ v_0"=37$ extends from 9 to 27 $a_0$ ($a_0$ denotes the Bohr radius) and the wave function of the resonant level $b^3\Pi$ $v_0'=43$ is located at much smaller internuclear distance, 7 $a_0$ to 11 $a_0$. As a result, the Franck-Condon factor is relatively small ($|\left\langle a v_0"=37|b v_0'=43\right\rangle|^2=1.16 \times 10^{-3}$).

\noindent
In the same Fig.~\ref{ch5:fig:3-levels} we also show
the wave function, in the Hund's case $a$ representation, of the spin-orbit-mixed vibrational level $0^+$ $v'=70$, which has  an energy close to the energy of the pure Hund's case $a$ resonant level $b\,v_0'=43$. One should notice the similarity between the vibrational component in the $b$ triplet state of the wave function $0^+$ $v'=70$ and the vibrational wave function of the pure $b\, v_0'=43$ level for $9.5 a_0 \le R\le 11 a_0$, that is in the $R$-range where the overlap of both wave functions is the largest. 

\noindent
The wave function of the $b\,v'=195$ level, strongly off-resonant with the studied laser pulse but connected to the $a\,v_0"$ level through a 'vertical' transition (the outer turning points of both wave functions are located at $R_{out}\sim 26$ $a_0$), is also reported in Fig.~\ref{ch5:fig:3-levels}. The corresponding Franck-Condon overlap, $|\left\langle a v_0"=37|b v'=195\right\rangle|^2=0.183$, is  much larger than that  one of the resonant transition.
\subsection{\label{ch5:subsec:pulse}Characteristics of the laser pulse}
\noindent
The laser pulse is assumed to have a Gaussian profile and to be Fourier-transform-limited, with a time-independent carrier frequency fixed to $\omega_L$. We do not consider chirped pulses because the mechanism of adiabatic population transfer occurring during  excitation with chirped pulses has been previously extensively analyzed and optimized \cite{cao1998, cao2000, luc2003, luc2009}. The motivation of the present work is to investigate a completely different excitation mechanism, resulting from the use of ultrashort unchirped pulses, and to interpret in detail its dynamics.
 
\noindent
The laser pulse is described by an electric field with an amplitude ${\cal{E}}(t)$ varying with time as:
\begin{eqnarray}
{\cal{E}}(t)&=&{\cal {E}}_0 f(t) \cos [\omega_Lt]={\underline{\cal{E}}}(t) + {\underline{\cal{E}}}^*(t)  \nonumber \\
            &=&\frac{{\cal{E}}_0}{2} f(t) \exp[i \omega_Lt] \, + \, \frac{{\cal{E}}_0}{2} f(t) \exp[-i \omega_Lt] \; ,
\label{ch5:eq:field}
\end{eqnarray}
\noindent
where ${\cal{E}}_0$ is the maximum amplitude and  ${\underline{\cal{E}}}(t)$ denotes the complex time-dependent amplitude. The Gaussian envelope $f(t)$, with maximum $f(t_P)=1$, is given by
\begin{equation}
f(t)=\exp\left[ -2 \ln2\left(\frac{t-t_P}{\tau_L}\right)^2\right] \; .
\label{ch5:eq:envelop}
\end{equation}

\noindent
The instantaneous intensity $I(t)$ of this pulse  illuminating an area $\sigma$, is equal to
\begin{align}
  \!\!\!I(t) &\!=\frac{E_{\mathrm{pulse}}}{\sigma\tau_L}
  \sqrt{\frac{4\ln2}{\pi}}
  \exp\!\!\left[ -4\ln2
      \left(\frac{t-t_P}{\tau_L}\right)^2 \right]=I_L[f(t)]^2 ,
\label{ch5:eq:pulse}
\end{align}
\noindent
where
 $I(t_P)=I_L=c \epsilon_0 {\cal{E}}_0^2/2$ ($c$ is the velocity of light, $\epsilon_0$  the vacuum permittivity). $I(t)$ has a full width at half maximum (FWHM) equal to $\tau_L$. The pulse duration and the energy $E_{\mathrm{pulse}}$ of the pulse satisfy:  
\begin{eqnarray}
 \label{ch5:eq:laser-energyP}
 \frac{\!E_{\mathrm{pulse}}}{\sigma} & = & \sqrt{\frac{\pi}{4\ln2}} \;I_L \; \tau_L \; . 
 \end{eqnarray}
 
\noindent
In the spectral  domain, the electric field ${\overline{\cal{E}}}(\omega)$ is obtained from the Fourier transform of the complex time-dependent electric field ${\underline{\cal{E}}}(t)$,
\begin{eqnarray}
{\overline{\cal{E}}}(\omega-\omega_L)&=&\frac{{\cal{E}}_0}{2\sqrt{2\pi}}\int_{-\infty}^{+\infty} 
f(t) \exp{[i\omega_L t]}\,\exp{[-i\omega t]  } dt  \nonumber \\
&=& \sqrt{\frac{ \ln{2}}{\delta\omega^2}} {\cal{E}}_0 \exp\left[-2 \ln2 (\frac{\omega - \omega_L}{\delta \omega})^2\right] \, \nonumber \\
&\times& \exp{[i(\omega_L-\omega) t_P]} . 
\label{ch5:eq:E-omega}
\end{eqnarray}

\noindent
For the pulse of duration $\sim 100$ fs considered here, the bandwith $\delta\omega ={4 \ln2}/{\tau_L}$, defined by the FWHM of $\left|{\overline{\cal{E}}}(\omega-\omega_L)\right|^2$, is of the order of $\sim 150$ cm$^{-1}$. 
\subsection{\label{ch5:subsec:TimeD-photEx}Photoexcitation dynamics: 'Wavepacket' and 'Level by Level' descriptions}

\noindent
To analyze the dynamics of the photoexcitation process (Eq.~(\ref{ch5:eq:excit})), we consider the time-dependent Schr\"{o}dinger equation describing the internuclear dynamics of the Rb and Cs atoms
\begin{equation}
 [\hat{H}_{mol}  - \vec{\mu} \cdot \vec{E}(t)] \Psi (t)=i\hbar\frac{\partial}{\partial t}\Psi (t) ,
  \label{ch5:eq:hamil-timeDep-ab}
\end{equation}
where $\hat{H}_{mol}$ denotes the molecular Hamiltonian in the Born-Oppenheimer approximation and where the coupling between the laser and the molecule, written in the dipole approximation, is expressed in terms of the dipole moment operator $\vec{\mu}$. The electric field of the laser pulse with polarization $\vec{e}_p$ reads  $\vec{E}(t)={\cal{E}}(t) \vec{e}_p$.

\noindent
In the excitation process, we focus on the redistribution of the population between the vibrational levels, disregarding  rotational components of the wavepackets $\Psi(t)$. This approximation is justified because the centrifugal energy is negligible and thereby vibrational wavepackets do not depend on  value of the total angular momentum $J$. All our calculations were carried out for a fixed value of $J$, $J=0$ and below we do not identify it explicitly. 

\noindent
In a simple  model restricted to the ground $g$ and excited $e$ electronic states, the two radial  components $\chi_g(R,t)$ and $\chi_e(R,t)$ of the wavepacket $\Psi(R,t)$   are solutions of the coupled system: 
\begin{eqnarray}
\label{ch5:eq:matricial2x2} 
 &&i\hbar\frac{\partial}{\partial t}\left(\begin{array}{c}
 \chi_{g}(R,t)\\
 \chi_{e}(R,t)
 \end{array}\right)\\
 &=&
 \left(\begin{array}{lc}
 -\frac{\hbar^2}{2\mu}\frac{\partial^2}{\partial R^2} + \overline{V}_g(R) & 
 W_{ge}(t) \cos[\omega_L t] \nonumber \\
 W_{eg}(t) \cos[\omega_L t] & 
 -\frac{\hbar^2}{2\mu}\frac{\partial^2}{\partial R^2} +  \overline{V}_e(R)    
 \end{array} \right) 
 \left( \begin{array}{c}
 \chi_{g}(R,t)\\
 \chi_{e}(R,t) 
 \end{array}\right)
\end{eqnarray}
where $\overline{V}_g(R)$ and $\overline{V}_e(R)$ denote the potentials in the ground and excited states. 
The coupling of the two electronic states can be written in terms of $W_{ge}(R,t)$ 
\begin{eqnarray}
W_{ge}(R,t) =-{D}_{ge}({R} ) {\cal {E}}_0 f(t) &=& -2W_L f(t), 
\label{ch5:eq:W} 
\end{eqnarray}
\noindent
where $D_{ge}(R)$ denotes the electronic dipole transition moment resulting from the integration of $\vec{\mu}$ over the electronic wave functions of the ground and excited electronic states. We disregard  the $R$-dependence of the electronic transition dipole, which is taken equal to its asymptotic value $\mathbf{D}$. Finally, $W_L={\mathbf{D}}{\cal{E}}_0/2$, the maximum strength of the coupling, is proportional to the square root of the maximum intensity $I_L$.

\noindent
The radial part of the wavepackets $\chi_g(R,t)$ (resp. $\chi_e(R,t)$) is a coherent superposition of the stationary vibrational wave functions,   eigenstates $\varphi_{g,v"}(R)$ with energy $E_{g,v"}$ (resp. $\varphi_{e,v'}(R)$ and $E_{e,v'}$) of the time-independent Schr\"{o}dinger equation involving the potential $\overline{V}_g(R)$ (resp. $\overline{V}_e(R)$). Numerically the radial dependences of all functions are described by using the Mapped Fourier Grid Method (MFGH) \cite{kokoouline1999,willner2004}. Let us emphasized that for a single potential, the eigenstates consist of bound levels and discretized scattering levels, which are automatically included in the decomposition of the wavepacket (see Appendix \ref{app:MFGH}). A spatial grid of length $L$ with $N$ mesh points  is used  for each potential yielding a quasi-complete set  of $N$ eigenfunctions (see Ref.~\cite{londono2011}). 

\noindent
Two methods are used  to solve the time-dependent Schr\"{o}dinger equation  in  the rotating wave approximation (RWA).  
\noindent 
The first method, the Wavepacket description, consists in determining directly  the vibrational wavepackets $\chi_g(R,t)$ and $\chi_e(R,t)$ created by the laser pulse on both electronic states $g$ and $e$. Studying the excitation from the vibrational level $a^3\Sigma^+ v_0"=37$ (Fig.~\ref{ch5:fig:3-levels}), the initial state is chosen to be  this initial vibrational level:  $\chi_{\mathrm{g}}(R,t=0)=\varphi_{a, v_0"}$ and $\chi_{\mathrm{e}}(R,t=0)=0$. Details on the numerical methods,  presented in Refs \cite{luc2003,luc2004}, are summarized in Appendix \ref{app:WP}. The time-dependent Schr\"{o}dinger equation is solved  by expanding the evolution operator in Chebyschev polynomials \cite{kosloff1994}. With the MFGH method being used to represent the radial dependence of the wavepackets, the WP method is a global approach which automatically incorporates contributions of the complete set of vibrational levels  $\varphi_{g,v"}(R)$ and $\varphi_{e,v'}(R)$ with $0\le v',v''\le N-1$ . 

\noindent
The second approach, the Level by Level description, analyzes the coupling by the laser pulse of some beforehand selected subsets of vibrational levels $\underline{g}_n$, $\underline{e}_m$, with $g_n$ and $e_m$ being numbers of levels in the ground and excited state vibrational subsets respectively. 

\noindent
The $(g_n+e_m)$ chosen levels 
result in the formation of the ground and excited wavepackets written, in the 'interaction representation' \cite{bookCohen2}, as:
\begin{eqnarray}
\underline{\chi}_g(R,t)&=&\sum_{v"\in \underline{g}_n} \underline{a}_{v"}(t) \exp\left[-i\frac{E_{g,v"}t}{\hbar} \right]\, \varphi_{g,v"}(R) ,\nonumber \\
\underline{\chi}_e(R,t)&=&\sum_{v'\in \underline{e}_m} \underline{b}_{v'}(t) \exp\left[-i\frac{E_{e,v'}t}{\hbar} \right]\, \varphi_{e,v'}(R)  , 
\label{ch5:eq:ampli-LbyL}
\end{eqnarray}
where the phase factor accounts for the 'free evolution' of the stationary vibrational levels. 
In the RWA approximation, the instantaneous probability amplitudes  $\underline{a}_{v"}(t)$ and $\underline{b}_{v'}(t)$ are determined by solving  a system of $(g_n+e_m)$  coupled first-order differential equations (Eq.~(\ref{ch5:eq:eq-coupl-RWA})) presented in the Appendix \ref{app:LbyL}. For the initial state of the system, the probability amplitude of the $a^3\Sigma^+\, v_0"=37$ level is set to unity: ${\underline{a}}_{v"}(t=0)=\delta(v",v_0")$ and ${\underline{b}}_{v'}(t=0)=0$ for all the considered $v'$ values. The relevant molecular structure data are the relative energies $\Delta^g_{v",v_0"}$ for the ground levels  (resp. $\Delta^e_{v',v_0'}$ for the excited levels) with respect to the resonant level $v_0"$ (resp. $v_0'$),
\begin{eqnarray}
\Delta^g_{v",v_0"} &=& E_{g,v"} - E_{g,v_0"} =\delta_{v_0',v"} , \nonumber \\
\Delta^e_{v',v_0'} &=& E_{e,v'} - E_{e,v_0'} = - \delta_{v',v_0"} ,
\label{ch5:eq:Delta-g-e}
\end{eqnarray}
\noindent
and the overlap integrals
\begin{eqnarray}
\langle v'|v" \rangle=\langle\varphi_{e,v'}|\varphi_{g,v"}\rangle .
\label{ch5:eq:overlap}
\end{eqnarray}

\noindent
Notice that because of the resonance condition, the energy spacings $\Delta^g_{v",v_0"}$ and $\Delta^e_{v',v_0'}$  may be expressed in terms of the detunings $\delta_{v',v"}$ (Eq.~(\ref{ch5:eq:delta})).

\noindent
The WP and LbyL methods are  compared in Appendix \ref{app:WP-LbyL}. The WP/MFGH approach allows one to expand the  wavepackets $\chi_g(R,t)$ and $\chi_e(R,t)$ over the complete set of $N$ vibrational levels of the $g$ and $e$ electronic states:
\begin{eqnarray}
\chi_g(R,t)&=&\sum_{v"=0,{N}-1} a_{v"}(t) \exp[-i\frac{E_{g,v"}}{\hbar} t]\, \varphi_{g,v"}(R) ,\nonumber \\
\chi_e(R,t)&=&\sum_{v'=0,{N}-1} b_{v'}(t) \exp[-i\frac{E_{e,v'}}{\hbar} t ]\, \varphi_{e,v'}(R)  . 
\label{ch5:eq:ampli-wp}
\end{eqnarray}
The evolution of the total population in the two electronic states may be found as
\begin{eqnarray}
 P_e(t)&=& 
 \left\langle \chi_e(R,t)|\chi_e(R,t)\right\rangle , \nonumber \\
  P_g(t)&= & 
  \left\langle \chi_g(R,t)|\chi_g(R,t)\right\rangle .
 \label{ch5:eq:populationWP}
\end{eqnarray} 

\noindent
More detailed information is provided by decomposing the wavepackets in the basis of unperturbed vibrational levels $v'$ or $v"$ of both electronic states $e$ or $g$,
\begin{eqnarray}
 P_{gv"}(t)&=& \left|\left\langle \varphi_{g,v"}(R)|\chi_{g}(R,t)\right\rangle\right|^2 =|a_{v"}(t)|^2 ,\nonumber \\
 P_{ev'}(t)&=& \left|\left\langle \varphi_{e,v'}(R)|\chi_{e}(R,t)\right\rangle\right|^2 =|b_{v'}(t)|^2,
 \label{ch5:eq:populationWP-descomp}
\end{eqnarray}
which gives the instantaneous population of each stationary vibrational level. 

For the LbyL approach, populations similar to those defined in Eqs.~(\ref{ch5:eq:populationWP}) and (\ref{ch5:eq:populationWP-descomp}) can be introduced.

\noindent
Naturally the LbyL approach is equivalent to the WP description if and only if the sets $\underline{g}_n$ and $\underline{e}_m$ encompass complete sets with $g_n=e_m=N$ levels, that is all bound levels and all levels of the discretized continua (Appendix \ref{app:MFGH}). We emphasize that the WP description automatically takes advantage of the completeness  of the set of eigenfunctions provided by the spatial representation of the Hamiltonian on a grid. Furthermore, the description of the dynamics does not depend on the choice of the grid, provided that a sufficiently wide domain of energy is covered by the eigenvalues obtained in the MFGH diagonalization of the Hamiltonian matrix.
\section{\label{ch5:sec:pi-pulse}Wave Packet description: from low field toward  ${\bm \pi}$-pulse }
\subsection{\label{ch5:subsec:pi-pulse-cond}${\bm \pi}$-pulse condition}

\noindent
Our goal is to find a pulse which yields a population transfer as large as possible from the initially populated vibrational level $a^3\Sigma^+$ $v_0"=37$ toward the vibrational level $b^3\Pi$ $v_0'=43$. As mentioned above, we consider only the case of an unchirped transform-limited Gaussian pulse, resonant with the transition $a^3\Sigma^+ v_0"=37 \rightarrow b^3\Pi v_0'=43$, with a duration in the femtosecond domain.
The chosen duration is $\tau_L=120$ fs, much smaller than the vibrational period $T_{g,v_0"}^{vib}=4\pi\hbar/(E_{g,(v_0"+1)} -E_{g,(v_0"-1)})\sim 22$ ps for the initial level $a^3\Sigma^+ v_0"=37$. It is only 6 times smaller than the vibrational period $T_{e,v_0'}^{vib}=0.72$ ps in the excited state. Consequently, in the excited electronic state, there are only 6 nearly-resonant levels lying within the bandwidth $\delta \omega = 122$cm$^{-1}=5.59 \times10^{-4}$ au of the pulse, the levels $41 \le v' \le 45$ with detuning $\delta_{v'v_0"}$ respectively equal to -92.0, -46.0, 0, +45.8, +91.5 cm$^{-1}$. 

\noindent
The pulse is characterized by the electric field amplitude ${\cal{E}}_0$ or, equivalently, by the pulse intensity $I_L$ or by the parameter $ W_L$ (Eq.~(\ref{ch5:eq:W})).  Given a pair of levels (say $v_0"$ and $v_0'$) we may  also introduce the accumulate pulse  area~\cite{bookTannor} as
\begin{equation}
 \Theta(t)= \mathbf{D}{\cal{E}}_0|\langle v_0'|v_0" \rangle |\int_{-\infty}^t f(t')dt' \, ,
\label{ch5:eq:area-pulse}
\end{equation}
where $\langle v_0'|v_0" \rangle$ denotes the overlap integral of the resonant transition (Eq.~(\ref{ch5:eq:overlap})).
The total pulse  area of a Gaussian pulse is  $$\Theta_F=\Theta(+\infty)=W_L \tau_L |\langle v_0'|v_0" \rangle | \sqrt{\frac{2\pi}{\ln2}} .$$

 
\noindent
In a two-level system, the angle $\Theta(t)$ fully determines the probability amplitudes of the lower level ${\underline{a}}(t)$ and of the resonantly-excited (i.e., when $\delta_{v_0',v_0"}=0$ ) level ${\underline{b}}(t)$ as \cite{bookTannor} 
\begin{eqnarray}
{\underline{a}}(t)= \cos\left[\frac{1}{2}{\Theta(t)}\right],   \; \;   {\underline{b}}(t)=i \sin\left[\frac{1}{2}\Theta(t)\right]. 
\label{ch5:eq:rabi-res-2levels}
\end{eqnarray}
 
\noindent
The  $\pi$-pulse for a resonantly-driven two-level system is defined as $\Theta_F=\pi$,  
\begin{eqnarray}
 \tau_L W_L \left| \left\langle v_0'|v_0"\right\rangle \right| = \sqrt{ \frac{ \pi \ln2}{2}} \;   \mathrm{or}  \; 4 \sqrt{\frac{2 \ln2}{\pi}} \Omega(t_P)=\delta \omega,
\label{ch5:eq:pi-pulse}
\end{eqnarray} 
where $\Omega(t_P)= W_L \left| \left\langle v_0'|v_0"\right\rangle \right|$ is the Rabi coupling (see Eq.~(\ref{ch5:eq:omega-t})) for the resonant transition  $v_0'\rightarrow v_0"$ at the pulse maximum  $t=t_P$ . 

\noindent
Accounting for  the overlap integral  {$\left| \left\langle v_0'|v_0"\right\rangle \right|= 0.03462  \; \rm{au}.$} and for the pulse duration $\tau_L=0.12$ ps$=4961.11$ au, the $\pi$-pulse condition is satisfied when $$W_L^\pi=\frac{1}{2}{\mathbf{D}}{\cal{E}}_0 =6.076\times 10^{-3} \; \rm{au} \;\,\rm{or}\,\;I_L=493GW/\rm{cm}^2.$$ This large value of intensity is  due to the small value of the overlap integral and to the short pulse duration.
%
\subsection{\label{ch5:subsec:low-field}Low field excitation}
\noindent
\begin{figure}
\begin{center}
\includegraphics[viewport=0 0 615 601, width=8cm, clip]{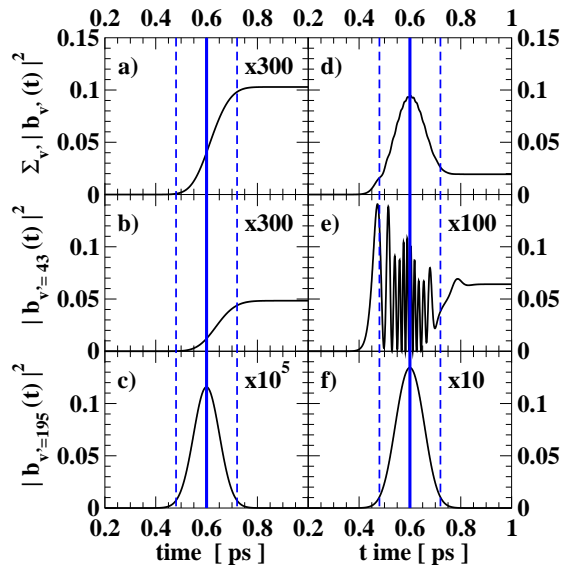}
\caption{\label{ch5:fig:w-h-field}(Color online) WP approach: variation with time (in ps) of the population in the excited electronic state $b^3\Pi$. Panels a), b), c): low field excitation $W_L=W_L^\pi/120$. Panels d), e), f):  high field excitation  $W_L=W_L^\pi$. Panels a) and d): total population $\sum_{v'=0}^{N-1}|b_{v'}(t)|^2$ in the excited electronic state. Panels b) and e):  population $|b_{v_0'=43}(t)|^2$ in the resonant level $v_0'=43$. Panels c) and f): population $|b_{v'=195}(t)|^2$ in the far from  resonance level  $v'=195$, corresponding to the `vertical transition'  defined in Fig.~\ref{ch5:fig:3-levels}.  Duration of the pulse $[t_P-\tau_L, t_P+\tau_L]$ (vertical  dashed blue [black] lines) maximum at $t=t_P$ (vertical thick continuous  blue [black] line). The populations have been multiplied by the factor indicated in the upper right  corner.}
\end{center}
\end{figure}

\noindent
We first consider a weak pulse, $W_L=5.\, 10^{-5}$ au, with a pulse area $\Theta_F= \pi/120$, corresponding to an intensity at the maximum of the pulse $I_L=I(t_P)=34\rm{MW}/$cm$^2$. The initial population in the $a^3\Sigma^+ v_0"=37$ level is set equal to unity. The evolution with time of the total population in the excited electronic state $b^3\Pi$ and in the resonant level $v_0'=43$ is reported in Fig.~\ref{ch5:fig:w-h-field}a,b. The considered populations increase monotonously during the pulse and the total transfer is very small (0.000343), with half population (0.000161) in the resonant level $v_0'=43$. For the $v'=42$ and $v'=44$ levels, which have a detuning with respect to the central laser frequency smaller than $\delta\omega/2$, the population at the end of the pulse is respectively 0.000076 and 0.000088. There is almost no population in the levels $v'\le40$ or $v'\ge46$.
\begin{figure}
\centering
\includegraphics[viewport=0 0 612 594, width=8cm, clip]{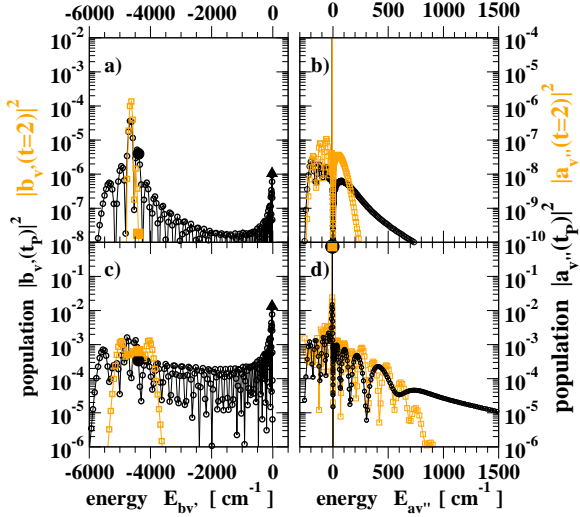} 
\caption{\label{ch5:fig:analysis}(Color online) WP approach: population $|b_{v'}(t)|^2$  in the levels $b^3\Pi v'$ (panels a) and c)) and population $|a_{v"}(t)|^2$ in the levels  $a^3\Sigma^+ v"$ (panels b) and d)), as a function of the  energy of the corresponding levels (in cm$^{-1}$) with respect to the  Rb(5$s$)Cs(6$p$) and Rb(5$s$)Cs(6$s$) dissociation limits respectively. The distributions of population are shown either at the maximum of the pulse $t=t_P=0.6$ ps (black dots) and after the end of the pulse $t=2$ ps (orange [gray] squares). Panels a) b): low field excitation $W_L=W_L^\pi/120$. Panels c) d): high field excitation $W_L=W_L^\pi$. The resonant level $v_0'$ is indicated by a large full black circle at $t=0.6$ ps and a large full orange  [gray] square at $t=2$ ps, the level $bv'=195$ is represented by a  full black triangle.}
\end{figure}
 
\noindent 
In the perturbative limit, the amplitude of population of the initial level is almost not modified during the pulse. After the end of the pulse, for $t\gg t_P+\tau_L$, the population of the level $v'$ in the excited electronic state is equal to:
\begin{eqnarray}
 |b_{v'}(t\rightarrow+\infty)|^2 & = & \frac{D^2{\cal{E}}_0^2}{4} \left\langle v'|v_0"\right\rangle ^2   \nonumber           \\
 &\times &  \left|\int_{-\infty}^{+\infty} \exp[-i \delta_{v',v_0"} t ]f(t) dt\right|^2 \nonumber \\
   & = & \frac{D^2}{4} \left\langle v'|v_0" \right\rangle ^2 8\pi \, \left|{\overline{\cal{E}}}( \frac{-\delta_{v',v_0"}}{\hbar})\right|^2  ,
   \label{ch5:eq:b-perturb}
\end{eqnarray}
\noindent
where $\delta_{v',v_0"}$ is the detuning of the excitation of the $ev'$ level from the $gv_0"$ level  and where $|{\overline{\cal{E}}}(\omega-\omega_L)|^2$ (Eq.~(\ref{ch5:eq:E-omega})) is the spectral density of the pulse.

\noindent
In this limit, the population transferred  from the level $gv_0"$ toward the level $ev'$ is proportional to the Franck-Condon factor $\left\langle v'|v_0"\right\rangle^2$ and to spectral density of the pulse at the excitation frequency \cite{shapiro2003}. As a result, for the weak perturbative pulses, only the 
nearly-resonant levels, such as $|\delta_{v',v_0"}|<\delta \omega$, are excited. 

\noindent 
The population distribution in the vibrational levels is presented in the left column of Fig.~\ref{ch5:fig:analysis} for the excited electronic state (panel a) and for the lowest electronic state (panel b), either at the maximum of the pulse ($t=t_P$) or after the end of the pulse ($t=2$ ps). The population of the excited vibrational levels $v'\sim 180-200$ always remains  smaller than that of the nearly-resonant levels $40\le v' \le 45$, and, at the end of the pulse, only these levels remain populated. In the low-field limit, the dynamics of the excitation process involves almost only the nearly-resonant levels (Figs. (\ref{ch5:fig:w-h-field}) and (\ref{ch5:fig:analysis})).
\subsection{\label{ch5:subsec:strongfield}Increasing the field strength}
\begin{figure}
\begin{center}
\includegraphics[viewport=0 0 581 551, width=7.5cm, clip]{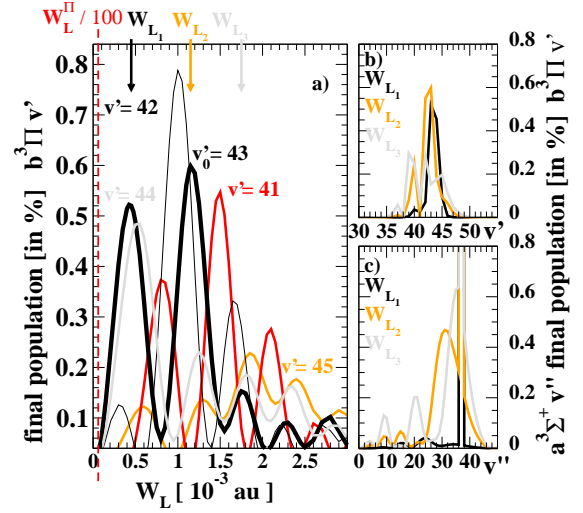}
\caption{\label{ch5:fig:c-croit}(Color online) WP approach. Panel a): variation of the population $|b_{v'}(t\rightarrow+\infty)|^2$ remaining after the pulse in the nearly-resonant excited levels $b^3\Pi v'$, as  function of the laser coupling $W_L$ (in units of $10^{-3}$ au); $v'=41$ (medium-thick red [black] line); $v'=42$ (thin  black line); $v'=v_0'=43$ (thick  black line); $v'=44$ (medium-thick   clear-gray line);  $v'=45$ (medium-thick  orange [gray]  line). The low field excitation $W_L^{\pi}/100$ is indicated by the thin dashed vertical red [black] line. The  couplings  $W_{L_1}=4.5 \times 10^{-4}$ au (black line), $W_{L_2}=1.15 \times 10^{-3}$ au (orange [gray] line) and $W_{L_3}=1.75 \times 10^{-3}$ au (clear-gray line) corresponding to the maxima of $|b_{v_0'=43}(t\rightarrow+\infty)|^2$ are indicated by the vertical arrows at the top  of the  panel. Right panels: couplings  $W_{L_i}\;i=1-3$. Panel b):  population 
 transferred to the  levels $b^3\Pi v'$ 
as function of  $v'$. Panel c): Population redistributed  in the levels $a^3\Sigma^+ v"$ 
 as function of  $v"$.}
\end{center}
\end{figure}

\noindent
Now we vary the laser coupling $W_L$ and explore the population $|b_{v'}(t\rightarrow+\infty)|^2$ transferred to the excited levels $b\,v'$ with $41 \le v'\le 45$. The results of our WP calculations are shown in Fig.~\ref{ch5:fig:c-croit}a.  In the low-field limit, the populations increase proportionally to $W_L^2$, and, as already noticed, only the levels $v'=42$, $43$ and $44$  are significantly populated. 
However, when the pulse area/intensity are increased, the population in the levels with $v'\le 41$ or $v' \ge 45$ becomes comparable to the population in the nearly-resonant levels. The population in the resonant level at the end of the pulse, $|b_{v_0'}(t\rightarrow+\infty)|^2$, first increases with increasing $W_L$ and reaches, for $W_L\sim W_L^\pi/14=0.000425$ au, a relatively small maximum, $0.0052 \sim 1/14^2$. This coupling corresponds for the resonant transition to an 'effective' pulse area of $\pi/14$, still in the low-field regime. As  $W_L$  is increased further, the  $|b_{v_0'}(t=+\infty)|^2$ oscillates with a period roughly equal to $\Delta W_L=0.0007$ au. Notice that as a function of $W_L$, the values of the population maxima decrease after two oscillations. 
This behavior strongly differs from what one would expect intuitively for the resonantly-excited two-level system [$ gv_0", \, ev_0'$]: in that case, the population would oscillate between the values of 0 and 1, with a period equal to $2W_L^\pi$, the value of 1 being reached at $W_L=W_L^\pi=6.076\times10^{-3}$ au.

\noindent
The population distribution among the levels of the excited $b^3\Pi$ and initial $a^3\Sigma^+$ electronic states after the pulse is presented in Fig.~\ref{ch5:fig:c-croit}b,c for three values of the coupling $W_L$. These couplings correspond to the first three maxima in the variation of $|b_{v'=43}(t\rightarrow+\infty)|^2$ as a function of $W_L$ (see the vertical arrows at the top of Fig.~\ref{ch5:fig:c-croit}a). For $W_{L_1}=4.5 \times 10^{-4}$ au, only three nearly-resonant levels are populated and no significant redistribution of population occurs in the $a v"$ levels. For $W_{L_2}=1.15 \times 10^{-3}$ au, more $b v'$ levels, with $39\le v'\le 46$, are populated and the population is recycled back to levels $a v"$ of the initial state with $25 \le v" \le 43$. For $W_{L_3}=1.75 \times 10^{-3}$ au, a still larger number of $a v"$ and $b v'$ levels is involved in the redistribution of population. 
\subsection{\label{ch5:subsec:pi-pulse-pi}${\bm \pi}$ pulse: resonant and far-from-resonance excitation}

\noindent
The time-evolution of the total population $\sum_{v'=0}^{N-1}|b_{v'}(t)|^2$ transferred to the excited electronic state $b^3\Pi$ during the excitation by a pulse with a large coupling strength $W_L^\pi$ is presented in Fig.~\ref{ch5:fig:w-h-field}d.  Population maximum (0.094) is attained at the maximum of the pulse $t=t_P$; it becomes smaller when the pulse intensity decreases. The final value, equal to 0.019, is much smaller than unity. The evolution of the population $|b_{v_0'}(t)|^2$ of the resonant level $v_0'=43$ is shown in Fig.~\ref{ch5:fig:w-h-field}e. This population does not increase monotonically, as one would expect for a $\pi$-pulse in a two-level system, but exhibits several ($\sim 11.5$) oscillations and the transfer is low (0.00064). A similar behavior is observed for the nearly-resonant levels $v'=42$ and $v'=44$ with final populations of  0.00043 and 0.00059, respectively. Figure \ref{ch5:fig:analysis} shows the population distribution  over various levels of the excited (panel c) and of the lowest (panel d) electronic states at two times  $t=t_P=0.6$ ps and  at $t=2$ ps. We find that at the end of the pulse a significant fraction of the population is transferred to a large number of strongly-bound $b^3\Pi\,v'$ levels, mainly to the  levels $26<v'<56$  with binding energies in the range of -5200 to  -3800 cm$^{-1}$. The most populated levels, $v'\sim 31$ and $v'\sim 51$, with respective detunings $\delta_{v',v_0"}=+560$ cm$^{-1}$ and $\delta_{v',v_0"}=-360$ cm$^{-1}$, have a population $\sim 0.0013$, equal to twice the population of the resonant level $v_0'=43$.  Population is also  redistributed within  bound and scattering levels of the ground   $a^3\Sigma^+\,v"$ state, in particular  within levels $32 \le v" \le 41$ (population $>$ 0.005). The difference in the energies of these levels with respect to the initially populated level $\Delta^g_{v",v_0"}=\delta_{v_0',v"}$ (Eq.~(\ref{ch5:eq:Delta-g-e})) lies in the range  $-4.2$ cm$^{-1} \le \delta_{v_0',v"} \le 11.3$ cm$^{-1}$. Only 76\% of the population remains in the initial $v_0"=37$ level. 

\noindent
At the maximum of the pulse, there are many levels of the excited electronic state, $v'\sim 180-200$, which have a population larger by a factor of at least 10 than the population in the nearly-resonant  $41\le v'\le 45 $ levels ($|\delta_{v',v_0"}| < 100$ cm$^{-1}$). These strongly-populated levels are such as $\delta_{v',v_0"} \sim 4300$ cm$^{-1}$, so they lie far outside the pulse bandwidth and correspond to highly-far-from-resonance excitations. Because of their high population during the pulse, these levels contribute significantly to the excitation dynamics. The time evolution of the population $|b_{v'=195}(t)|^2$ of the $b^3\Pi v'=195$ level, is reported in Fig.~\ref{ch5:fig:w-h-field}f. This is the most populated level in the excited electronic potential with a population reaching 0.0134 at the maximum of the pulse. The time-dependence of this population follows that of  the envelope of the pulse intensity, $|f(t)|^2$  (Eq.~(\ref{ch5:eq:envelop})). Note that in the low-field case (Fig.~\ref{ch5:fig:w-h-field}c), the population of the $v'=195$ level is always negligible ($<10^{-6}$).
 
\noindent 
It is to be emphasized that 
 this behavior can not be explained as Rabi cycling, contrarily to what could be intuitively expected considering the large value of the  instantaneous Rabi coupling ${\underline{\Omega}}(v',v_0",t)$ arising from the large value of the detuning $\delta_{v'=195,v_0"}$. We recall the definition of the instantaneous Rabi coupling at time $t$ for non-resonant transition $a\,v" \rightarrow b\,v'$:
\begin{equation}
{\underline{\Omega}}(v',v",t)=\frac{1}{2} \sqrt{ |\delta_{v',v"}|^2 + 4|f(t)  \left\langle v'|v"\right\rangle  W_L |^2 },
\label{ch5:eq:omega-t}
\end{equation}
\noindent
where $\left\langle v'|v"\right\rangle ^2$ denotes the Franck-Condon factor.
\noindent
The $b^3\Pi v'=195$ level is the level of the excited electronic state possessing the largest population at the maximum of the pulse. This can be understood by reminding that this level is excited from the initial level $a v_0"$ through a vertical transition (see Sec. \ref{ch5:subsec:reaction}).

\noindent
The importance, in the strong field regime, of off-resonant excitation of levels strongly favored by high Franck-Condon factors but lying energetically above the spectral bandwith of the pulse has been experimentally observed in the photoassociation of ultracold atoms with shaped femtosecond pulses \cite{salzmann2008,mccabe2009}.
\section{\label{ch5:sec:LbyLanalysis}Analysis of the $\pi$-pulse dynamics: Level by Level description} 
\begin{figure}
\centering
\includegraphics[viewport=0 0 568 590, width=8cm, clip]{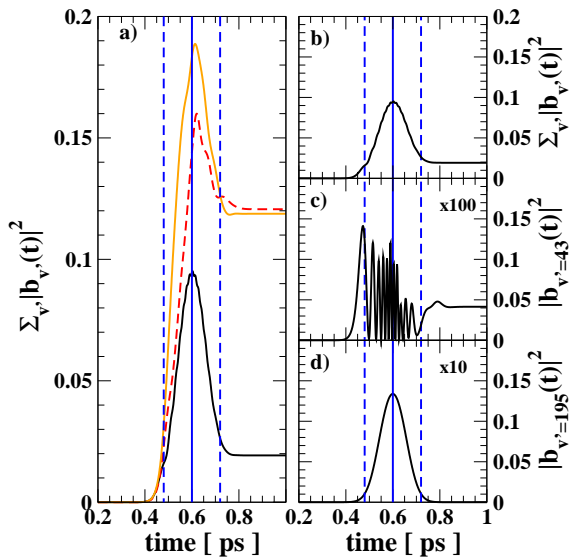}
\caption{\label{ch5:fig:a-b-levels-d}(Color online) LbyL description with three different basis sets of levels   in  the lower ($v"$) or the excited ($v'$) electronic states; these sets, labeled $A$, $B$ and $C$, are defined in the text:  variation of the population in the excited  b$^3\Pi v'$ levels as a function of time (in ps). The pulse (vertical continuous and dashed blue [black] lines) satisfies the $\pi$-pulse condition.  Panel a): total population for set $A$ (continuous orange [clear-gray] line) , $B$ (dashed red [dark-gray] line) and $C$ (continuous black line).   Panels b), c) and d): 'optimal' basis set $C$ $[v"=0-204, \, v'=0-218]$ reproducing the results of the WP method (Fig.~\ref{ch5:fig:w-h-field}); b): total population in the  bound levels $v'=0-218$; c): population in the resonant level  $v_0'=43$; d): population in the off-resonant level $b^3\Pi$ $v'=195$, corresponding to the vertical transition. Some populations have been multiplied by the factor indicated in the upper right corner.}
\end{figure}

\noindent
The WP results demonstrate that, for the high coupling strength $W_L^\pi$, the dynamics of the excitation process involves a large number of vibrational levels, both in the ground $a^3\Sigma^+$ and in the excited $b^3\Pi$ electronic states. To better understand the dynamics of the  population of these levels, we performed LbyL calculations, with various  subsets  $\underline{g}_n$ and $\underline{e}_m$ of bound and quasi-continuum (scattering) levels.  
These subsets are simply denoted as: $[v_2",...v_{g_n}", \, v_1',v_2',...v_{e_m}']$. 
\subsection{\label{ch5:subsec:levels} Levels involved in the dynamics}
\subsubsection{\label{ch5:subsubsec:large-basis} LbyL basis set reproducing the WP dynamics}

\noindent
In the first step we try to reproduce, by optimizing the restricted LbyL basis set, the time-evolution of the total population transferred to the excited electronic state by $\pi$-pulse  ($W_L=W_L^\pi$). 
Some representative  results  are displayed  in the left column of Fig.~\ref{ch5:fig:a-b-levels-d}, where the following  basis sets are considered: set $A$: $[v"=30-50, \,v'=0-218]$, set $B$: $[v"=20-50, \, v'=30-50,190-200]$ and set $C$: $[v"=0-204, \,v'=0-218]$. These levels  are either bound or discretized scattering vibrational levels  in the $a^3\Sigma^+$ or  $b^3\Pi$ electronic states. Let us remark, that, with the mesh grid  used in the MFGH approach, only a small energy range ($0<E<0.01$ cm$^{-1}$) is described by 'physical' scattering levels (see Appendix \ref{app:MFGH}).

\noindent
The relatively large set $B$  includes, in the lower state, bound levels lying close to the initial one, $v_0"=37$, and, in the excited state, levels located in the vicinity of the resonantly excited $v_0'=43$ level or in the vicinity of the far-from-resonance $v'=195$ level corresponding to the vertical transition. For this set, the total population at the maximum of the pulse $t=t_P$ is larger by a factor 2 than the population obtained by using the WP approach. At the end of the pulse, a too large population ($\sim 0.12$) remains in the excited state. For the set $A$, which includes all the bound levels in the excited state and, in the lower state, a smaller number of levels located in the vicinity of the initially populated one, a similar behavior is obtained, yielding the same final population transfer, but a slightly smaller maximum value at $t\sim t_P$. 
  
\noindent
To reproduce in the LbyL approach the results obtained in the WP approach, we have found that it is necessary to employ the set $C$ which includes all bound vibrational levels in the excited state and a very large number of levels (205) in the lower state, {\it{i.e.}} all bound levels ($0 \le v" \le 48$) and discretized scattering levels in a large energy range, with an energy up to $1100$ cm$^{-1}$, described by physical or unphysical levels \cite{londono2011}. In this LbyL calculation, the time-evolution of the total population in the excited state reproduces the one from the WP approach, in particular the low value of the population ($\sim 0.025$) transferred at the end of the pulse. Furthermore, the time-dependence of the populations in the resonant level $v_0'=43$ or in the level $v'=195$  and also the variation of the total population in the bound $b^3\Pi$ levels, represented in the right panel of Fig.~\ref{ch5:fig:a-b-levels-d}, reproduce perfectly the variations calculated directly in the WP approach (Fig.~\ref{ch5:fig:w-h-field}). In the following, the set $C$ is called the 'optimal' LbyL basis set.

\noindent
The wide energy range covered by the levels involved in the dynamics is not negligible compared to the frequency of the pulse $\hbar \omega_L\sim 7000$ cm$^{-1}$. Therefore the validity of the RWA approximation is questionable. Indeed, for the pairs of levels $E_{e,v'},\, E_{g,v"}$ included in the basis set, the frequencies of the 'rotating' contributions $[\frac{E_{e,v'}- E_{g,v"}}{\hbar}-\omega_L]$ are not always negligible compared to the frequencies of the neglected 'counter-rotating' contributions $[\frac{E_{e,v'}- E_{g,v"}}{\hbar}+\omega_L]$ (Appendix \ref{app:LbyL}). Further investigation would be needed to check that the introduction of the counter-rotating terms does not change the main conclusions of the present analysis. 
\begin{figure}[ht]
\centering
\includegraphics[viewport=0 0 710 529, width=8cm, clip]{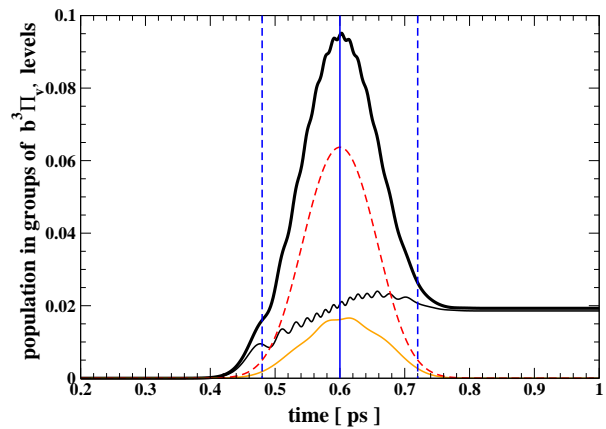}
\caption{\label{ch5:fig:3-groups}(Color online) LbyL description, optimal basis set $C$: variation of the population in the excited  b$^3\Pi v'$ levels as a function of time (in ps). The laser pulse (vertical continuous and dashed blue [black] lines) satisfies the $\pi$-pulse condition. Total population of the excited bound levels $v'=0-218$ (thick continuous black line). Non-adiabatic evolution for the population in the levels $v'=13-60$ (thin continuous black line) resulting in a non vanishing final population transfer. Adiabatic evolution, following the variation of the pulse intensity, for the far-from-resonance levels $v'=167-218$ (thin dashed red [black] line) located around the vertical transition $v'\sim 195$ and the most populated at the maximum of the pulse  or for the intermediate group of levels $v'=61-166$ (thin continuous orange [gray] line). }
\end{figure}
\subsubsection{\label{ch5:subsubsec:2-types}Two types of dynamics in the excited electronic state}

\noindent
To go further in the analysis of the dynamics, we separate the excited levels of the optimal set $C$ into two different groups, according to the time-evolution of their individual population $|\underline{b}_{v'}(t)|^2$.

\noindent
For levels $0\le v' \le 60$ with a detuning varying in the range $2200$ cm$^{-1} \ge \delta_{v' v_0"} \ge -790$ cm$^{-1}$,  the dynamics of population is very similar to that of the resonant level $v_0'=43$. During the pulse the population $|\underline{b}_{v'}(t)|^2$ exhibits a small number of oscillations of a relatively small amplitude and  some population remains in these levels after the pulse. The sum of the population in this group of levels grows almost monotonically during the pulse and reaches the final value $\sim 0.02$ (Fig.~\ref{ch5:fig:3-groups}).

\noindent
As we move further off-resonant and consider bound levels $61 \le v' \le 218$, we find that the evolution of the population is similar to that of the level $v'=195$, i.e., traces time-variation of the pulse intensity $\propto[{\cal{E}}(t)]^2$. The total population transferred  at  the maximum of the pulse $t=t_P$ is very high $\sim 0.08$ and it is larger than the population present in the group of  levels close to the resonance. Yet no population remains  after the end of the pulse.

\noindent
Thus it appears that two types of dynamics are observed for the levels of the excited electronic state. Levels with a not-too-large detuning remain populated after the laser pulse. Taking into account that the pulse is symmetrical, Gaussian and unchirped, their evolution is necessarily non-adiabatic. Conversely, levels corresponding to highly-off-resonant excitation possess the largest population at the maximum of the pulse, but they do not retain their population after the pulse: such dynamics has thus a quasi-adiabatic character. Below we present a qualitative description  which emphasizes a relation between detuning and adiabaticity.
\subsection{\label{ch5:subsec:two-level-s}Adiabaticity}
\subsubsection{\label{ch5:subsubsec:adiab-N-level}Introduction} 

\noindent
Adiabaticity of the evolution of a system is naturally expressed in the basis of instantaneous eigenvectors of the Hamiltonian, the so-called adiabatic basic \cite{bookTannor,bookMessiah} (see Appendix \ref{app:multilevel}). For a system with more than two levels, there is no general way to construct the instantaneous adiabatic basis and thus no general expression of the adiabatic theorem \cite{bookMessiah}. In fact, the relationship between adiabaticity, detuning, laser width and coupling strength can be perfectly illustrated in the case of  a two-level system $[v",\,v']$, where the instantaneous adiabatic levels can be explicitly constructed. The unperturbed vibrational levels $|g\rangle\equiv|v"\rangle$ and $|e\rangle\equiv|v'\rangle$, define the diabatic basis (see Appendix \ref{app:multilevel}). The time-dependent wave function can be decomposed on the diabatic levels $\Psi(t)=a(t)|g\rangle+b(t)|e\rangle$. We assume that only the level $|g\rangle$ is initially populated. The levels are coupled by a Gaussian pulse ${\cal{E}}_0f(t)$, with bandwidth $\delta\omega=5.59\,10^{-4}$ au, as described in Section \ref{ch5:subsec:pulse}. In the RWA approximation, the time-dependent coupling is $\Omega(t)= {\mathbf{D}} \times {\cal {E}}_0 \left\langle v'|v"\right\rangle f(t)/2$.

\noindent
Below we study six different cases, labeled $a$ to $f$; these differ by overlap integrals and  detunings (see Table \ref{table2}). For the overlap integral, we choose values corresponding either to the resonant transition  $\left\langle v_0'=43|v_0"=37\right\rangle$ (systems $a$ to $d$) or to the vertical transition $\left\langle v'=195|v_0"=37\right\rangle=12.3\,\left\langle v_0'|v_0"\right\rangle$ (systems $e$ and $f$). The amplitude of the electric field ${\cal {E}}_0$ and the dipole transition moment $\mathbf{D}$ are chosen such as, for $\left\langle v'|v"\right\rangle=\left\langle v_0'|v_0"\right\rangle$, the $\pi$ pulse condition, or $W_L=W_L^\pi$, is satisfied except for the cases $e$ and $f$, where $\Theta_F\sim 12.3\,\pi$. Therefore the maximum coupling is either smaller, $\Omega(t_P)=\delta\omega/2.66$ (cases $a$ to $d$), or larger, $\Omega(t_P)=4.63\,\delta\omega$ (cases $e,f$), than the pulse bandwidth.

\begin{table}
\caption{\label{table2}Different two-level systems $[v",\,v']$ considered here. $\delta$ is the detuning of the pulse, $\langle v'|v"\rangle$ is the overlap integral between the wave functions of the two levels and $\Omega(t_P)$ is the coupling at the maximum of pulse. The pulse has a duration $\tau_L=0.12$ ps and a bandwidth $\delta\omega=5.59\times 10^{-4}$ au.}
{\footnotesize {
\begin{tabular}{|c|c|c|c|c|}
\hline
system & excitation & $\delta$ [au] & $\langle v'|v"\rangle$  & $\Omega(t_P)$ [au]\\
\hline
\hline
$a$ & quasi  &     $10^{-8}$ & $0.03462$ & $2.10\times 10^{-4}$  \\
    & resonant &            &            &                     \\
\hline
$b$ & nearly & $10^{-4}$ &  $0.03462$ & $2.10\times 10^{-4}$  \\
    &resonant &         &             &                      \\
\hline
$c$ & out of & $10^{-3}$ &  $0.03462$ & $2.10\times 10^{-4}$  \\
    & resonance &        &            &                      \\
\hline
$d$ & out of & $2\times 10^{-3}$  & $0.03462$ &   $2.10\times 10^{-4}$  \\
    & resonance &              &             &                        \\
    \hline
$e$ & quasi &  $10^{-8}$  & $0.427364$  &  $2.59\times 10^{-3}$ \\
    & resonant &       &                & \\
\hline
$f$ & far from & $0.0198$ & $0.427364$  &  $2.59\times 10^{-3}$ \\
    & resonance &          &            & \\
\hline
\end{tabular}
}}
\end{table}

\noindent
We consider four classes of detunings. When the detuning satisfies $|\delta| \ll \delta \omega$, the systems are called `quasi-resonant' (cases $a$ and $e$). Systems where the detuning is  $|\delta| < \delta \omega$ are called `nearly-resonant' (case $b$). Larger detunings such as $|\delta| > \delta \omega$ correspond to `off-resonant' excitation (cases $c$ and $d$). 'Far-from-resonance' excitation, such as $|\delta| \gg \delta \omega$ is represented by the case $f$, where the detuning $\delta=35.4\,\delta\omega$ is the one of the vertical transition $v_0" \rightarrow v'=195$.
\subsubsection{\label{ch5:subsubsec:diab-basis} Two-level system: diabatic basis}

\begin{figure*}
\centering
\includegraphics[viewport=0 0 612 610, width=15cm, clip]{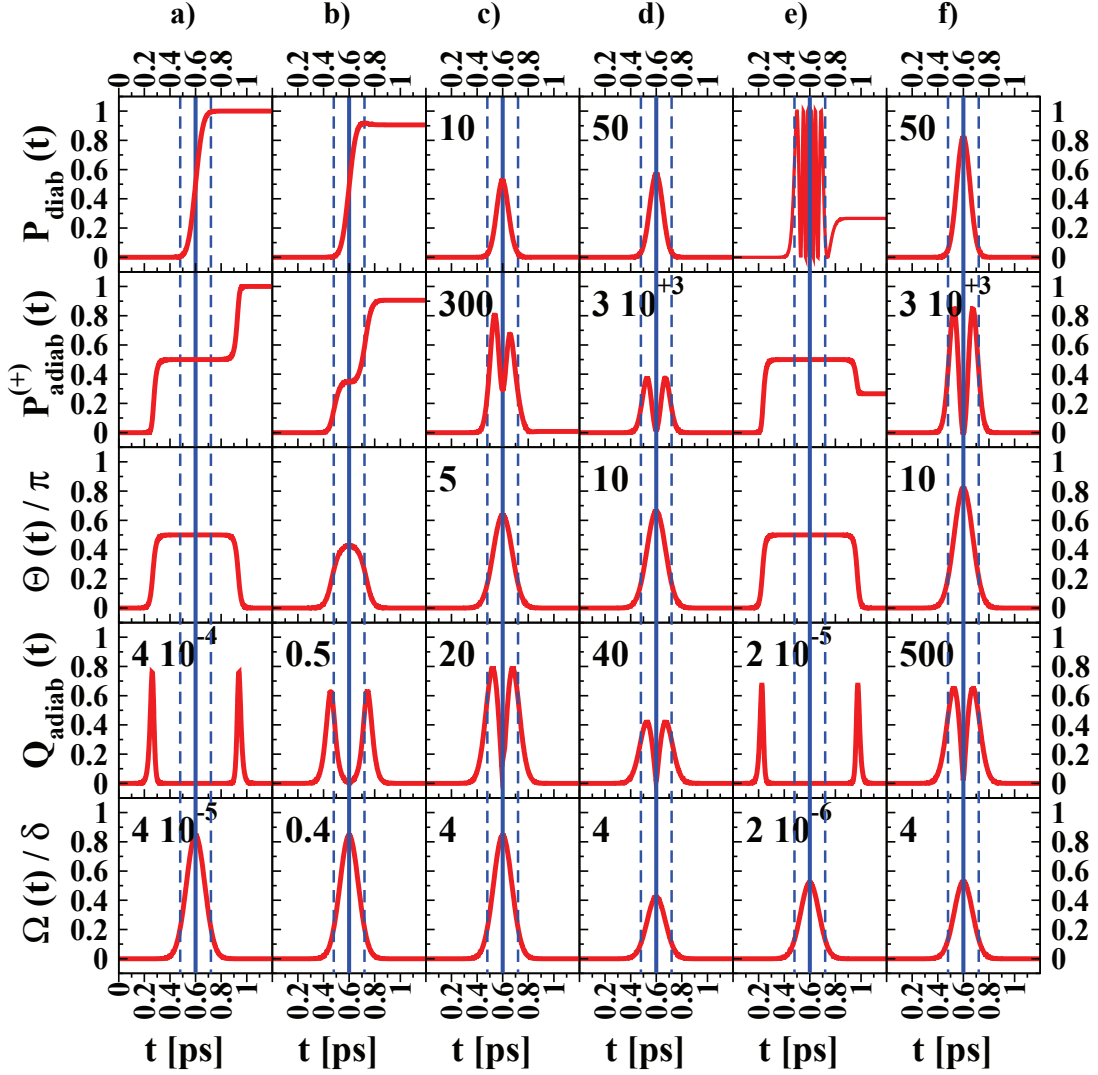}
\caption{\label{ch5:fig:adiab}(Color online) Two-level system: analysis, as a function of time (in ps), of the dynamics  for the different  cases $a$ to $f$ described in Table \ref{table2}. The maximum  and the duration of the pulse are indicated by the  vertical continuous and dashed blue [black] lines. 
First row:  population $P_{diab}(t)$ in the diabatic excited level $[e\rangle$, the initial population lying in the lower diabatic level $[g\rangle$. Second row: variation with time of the population $P_{adiab}^{(+)}(t)$ in the adiabatic level $\Psi_+(t)$, the initial population lying in the adiabatic level $\Psi_-(t=0)\equiv | g\rangle$. Third row:  rotation angle $\theta(t)/\pi$ (Eq.~(\ref{ch5:eq:angle-rotation})) defining the instantaneous adiabatic levels $\Psi_\pm(t)$. Fourth row:  parameter $Q_{adiab}(t)$ (Eq.~(\ref{ch5:eq:Q-adiab})) characterizing the adiabaticy  of the instantaneous transfer. Fifth row:  pulse parameter $\Omega(t)/\delta$. The quantities have been multiplied by the factor indicated in the upper left  corner.}
\end{figure*}

\noindent
For the six considered cases, the time-variation of the population $P_{diab}(t)=|b(t)|^2$ of the excited diabatic level $|e\rangle$ is calculated by solving the coupled system Eq.~(\ref{ch5:eq:2-level-diabatic}). The results are presented in the first row of Fig.~\ref{ch5:fig:adiab}. For quasi-resonant systems, the time-dependence of the population exhibits oscillations very similar to the Rabi oscillations of a resonantly excited two-level 
system. In the case $a$, the population, initially in the ground level, is transferred continuously to the excited level during the pulse. In the quasi-resonant case $e$, the population oscillates  more than 6 times from 0 to 1  between the ground and the excited levels, in agreement with the increase of the pulse area; at the end of the pulse 26.62\% of the population remains in the excited level. For the nearly-resonant case $b$, almost all the population, up to 91.6\%, is transferred monotonously to the excited level. 

\noindent 
For the off-resonant cases $c$ and $d$ and for the far-from-resonance case $f$, the time-evolution of the population in the excited level follows a Gaussian evolution similar to that of the pulse intensity $\propto [f(t)]^2$.  After the end of the pulse, the total population  returns  to the initial level. When the maximum value of the coupling is small  (for approximately $\Omega(t_P)/|\delta|<3/10$), the perturbative limit is valid and $P_{diab}(t)\sim [\Omega(t)/\delta]^2$. This occurs in the systems $c$, $d$ and $f$, where $\Omega(t_P)/\delta=0.21$, $0.10$ and $0.13$ respectively; this  can be also verified by comparing in Figs.~\ref{ch5:fig:adiab}c,d,f the first and fifth rows. The maximum value of the population in the excited diabatic level is small  and varies  proportionally to $|\langle v_0'|v_0" \rangle/\delta|^2$.
\subsubsection{\label{ch5:subsubsec:adiab-basis}Two-level system: adiabatic basis}

\noindent
To resume the analysis of the adiabaticity of the population transfer, we introduce now the adiabatic basis, made of the instantaneous eigenstates $|\Psi_\pm(t)\rangle$ of the Hamiltonian (see Appendix \ref{app:subsec:diab-2-level}). The time-dependent wave function is decomposed on the adiabatic levels $\Psi(t)=\exp({-i\frac{\delta}{2}t})\,[\alpha(t)\Psi_-(t)+\beta(t)\Psi_+(t)]$. For a fully adiabatic process, the populations $|\alpha(t)|^2$ and $|\beta(t)|^2$ remain constant during the pulse. In particular, if the system is initially in the adiabatic level $|\Psi_-(t=0)\rangle\equiv|g\rangle$, it remains in the instantaneous adiabatic level $|\Psi_-(t)\rangle$ during the pulse and  is in the level $|\Psi_-(t\rightarrow+\infty)\rangle$ after the end of the pulse. For a non-adiabatic process, the population of the adiabatic levels varies, and the stronger the non-adiabatic instantaneous population transfer is, the more rapid and important the changes of the instantaneous adiabatic populations. In our study, the detuning $\delta$ is fixed and only the coupling $\Omega(t)$ varies with time. Therefore $|\Psi_-(t\rightarrow+\infty)\rangle\equiv |g\rangle$. Simply, unless there is a significant non-adiabaticity, there is no population transfer. Conversely, a measure of the global non-adiabaticity of the process is related to the  population transfer to the excited level after the pulse. 

\noindent
The time-dependence  of the population $P_{adiab}^{(+)}(t)=|\beta(t)|^2$ in the adiabatic level $|\Psi_+(t)\rangle$ is obtained by solving the coupled system Eq.~(\ref{ch5:eq:2-level-adiabatic}), assuming that the population is initially in the adiabatic level $|\Psi_-(t=0)\rangle\equiv|g\rangle$ ($|\alpha(t=0)|=1$). The results are drawn in the second row of Fig.~\ref{ch5:fig:adiab}.  

\noindent
To characterize the adiabatic character of the instantaneous population transfer, we introduce the parameter $Q_{adiab}(t)$ deduced from the time-dependent Schr\"{o}dinger equation in the adiabatic basis (Eq.~(\ref{ch5:eq:schrodinger-adiabatic-d})):
\begin{eqnarray}
Q_{adiab}(t)&= &\frac{1}{2}\hbar\dot{\theta}(t)/\left(E_+(t) - E_-(t)\right) \nonumber \\
&=& \frac{1}{2}\frac{\delta \dot{\Omega}(t)}{[\delta^2 + 4(\Omega(t))^2]^{3/2}} \, .
\label{ch5:eq:Q-adiab}
\end{eqnarray} 
In these equations, $E_\pm(t)$ (Eq.~(\ref{ch5:eq:adiabatic-e})) denote the energies of the instantaneous adiabatic levels and $\theta(t)$ (Eq.~(\ref{ch5:eq:unitaryMatrix})) is the rotation angle occuring in the unitary matrix defining the adiabatic levels. Here and hereafter, the dot indicates the time derivative.  

\noindent
A strongly non-adiabatic instantaneous transfer corresponds to a high $Q_{adiab}(t)$ value. From Eq.~(\ref{ch5:eq:Q-adiab}), one can deduce that non-adiabaticity occurs for a small detuning $|\delta|$, for a small coupling strength $\Omega(t)$, {\it i.e.} at the beginning and at the end of the pulse or for a pulse with low intensity, and also when the rotation angle $\theta(t)$ varies rapidly. 

\noindent
The  time-dependences of the rotation angle $\theta(t)/\pi$ and of the parameter $Q_{adiab}(t)$ are shown respectively in the third and fourth rows of Fig.~\ref{ch5:fig:adiab}.  
\subsubsection{\label{ch5:subsubsec:adiab-resonant}Two-level system: from quasi-resonant to far-from-resonance excitation}

\noindent
The described  two-level model is useful for understanding the role of adiabaticity for both  resonant to far-from-resonance excitations.
 
\noindent
For strictly resonant excitation, $\delta=0$, and the rotation angle (Eq.~(\ref{ch5:eq:angle-rotation})) is equal to $\theta(t)=\frac{\pi}{2}$ at every time; then the initial wave function $|g\rangle$ corresponds to an equal mix of the two adiabatic levels $|g\rangle={1}/{\sqrt{2}}[|\Psi_-(t=0)\rangle-|\Psi_+(t=0)\rangle]$. During the evolution, there is no non-adiabatic coupling ($\stackrel{\cdot}{\theta}(t)=0$) and no change in the population of the adiabatic levels. The population oscillates between the ground and excited levels at the instantaneous Rabi frequency ${\underline{\Omega}}(v', v",t)/\hbar$.

\noindent  
For quasi-resonant excitation, with a very small detuning $|\delta| \ll \Omega(t_P)$, as for example $\delta=10^{-8}$ au   (cases $a$ and $e$), the rotation angle $\theta(t)$ is almost equal to zero at the beginning and at the end of the pulse, when $\Omega(t)\ll |\delta|$ (Fig.~\ref{ch5:fig:adiab}). Conversely, when $\Omega(t) \gg |\delta|$, the rotation angle remains constant and equal to $\theta=\frac{\pi}{2}$. For $t\sim t_{na_\pm}$, with $2|\Omega(t_{na_\pm})|=|\delta|$ (or $\theta(t_{na_\pm})=\frac{\pi}{4}$), the rotation angle  changes rapidly. Two sets of nearly adiabatic levels can be introduced, $\Psi^0_\pm(t)$,  valid  at the beginning $t<t_{na_-}$ or at the end of the pulse $t>t_{na_+}$ and  $\Psi^P_\pm(t)$, valid during the pulse $t_{na_-}<t<t_{na_+}$. During these three time intervals, the evolution is completely adiabatic.
For $ t \sim t_{na_\pm}=t_P \pm \tau_L[\frac{1}{2\ln{2}}\ln(\frac{{\mathbf{D}}{\cal{E}}_0|<v_0'|v_0"\rangle|}{\delta})]^\frac{1}{2}$,  the  
non-adiabatic couplings $Q_{adiab}(t_{na_{\pm}})\sim \frac{\sqrt{\ln2}}{2\delta\tau_L} \sqrt{\ln[{\frac{2 \Omega(t_P)}{|\delta|}}]}$ are huge, of the order of $3\times 10^4$ in the quoted examples. Therefore, strong instantaneous population transfer between the adiabatic levels occurs only around $t_{na_\pm}$. The value $|t_{na_\pm}-t_P|\sim 0.34$ ps is much larger than the pulse duration $\tau_L= 0.12$ ps. The population transfer occurs thus in the wings of the Gaussian pulse, at the turn-on and turn-off of the pulse, when the laser intensity is almost negligible. For $t<t_{na_-}$, the population remains in the lowest adiabatic level, described by the wave function $\Psi(t)=|g\rangle  \exp{[-\frac{i}{\hbar}E_gt]}$. If one sets $E_g=0$, then there is no change in the phase of this wave function. For $t_{na_-}<t<t_{na_+}$, $\theta(t)=\frac{\pi}{2}$ and the  adiabatic levels $\Psi_\pm^P(t)$ correspond at each time to an equal mix of both diabatic levels with a phase varying with time (Eq.~(\ref{ch5:eq:adiabaticity-wf})). These adiabatic levels evolve as
\begin{eqnarray}
|\Psi_\pm^P(t)\rangle&=&\frac{1}{\sqrt{2}}\left[\mp |g\rangle + |e\rangle\right] \exp[-\frac{i}{\hbar}\int_{t_{na_-}}^tE_\pm(t')dt'].\nonumber 
\label{ch5:eq:psi-adiab-pulse}
\end{eqnarray}
For $t_{na-}<t<t_{na+}$, the wave function $\Psi(t)$ can be decomposed on these states, with amplitudes $\alpha^P(t)$ and $\beta^P(t)$. The absolute values of these amplitudes remain constant (see row 2 in Fig.~\ref{ch5:fig:adiab}a,e). These populations can be estimated in the sudden approximation \cite{bookMessiah}, by projecting the adiabatic wavefunction $|\Psi^0_-(t_{na_-})\rangle$, valid just before $t_{na_-}$, on the adiabatic functions $\Psi^P_\pm(t_{na_-})$, valid just after $t_{na_-}$. In this way, one obtains $|\alpha^P(t)|=|\beta^P(t)|=\frac{1}{\sqrt{2}}$, just like for the strictly resonant excitation. The Rabi oscillations occurring at quasi resonance (Fig.~\ref{ch5:fig:adiab}e) in the population of the excited diabatic levels $|<\Psi(t)|e\rangle|^2$, for $t_{na_-}<t<t_{na_+}$, result from a beating effect in the coherent superposition of the $\Psi^P_\pm(t)$ adiabatic levels. At $t=t_{na_+}$, a strong non-adiabatic coupling occurs again during a very short time. The sudden approximation allows one again to obtain the value of the final population of the diabatic levels $|g\rangle$ and $|e\rangle$ after the end of the pulse, in terms of the quantity $\Theta'=\int_{t_{na_-}}^{t_{na_+}}[E_+(t)-E_-(t)] dt$, which, in the limit $\delta\rightarrow 0$, is equal to the pulse area $\Theta(+\infty)$.    

\noindent
With increasing $\delta$,  $|t_{na_\pm}-t_P|$ and  $Q_{adiab}(t_{na_\pm} )$ decrease, being equal in the nearly-resonant system $b$ to $|t_{na_\pm}-t_P|\sim 0.15$ ps and $Q_{adiab}(t_{na_\pm} )\sim 1.2$. Non-adiabatic transfer of population from the adiabatic level $\Psi_-(t)$ to the  adiabatic level $\Psi_+(t)$ occurs in two steps around $t_{na_\pm}$, but with a less-pronounced sudden character. 

\noindent
The maximum value of  $\theta(t_P)$ decreases when $|\delta|$ increases, and for a detuning such as $|\delta|>{2}\Omega(t_P)$ or $\theta(t_P)<\frac{\pi}{4}$, as in cases $c$, $d$ and $f$, the times $t_{na_\pm}$ do not exist. The maxima of the parameter $Q_{adiab}(t)$ become smaller and appear during the pulse at $t_\pm$ ($|t_\pm-t_P|<\tau_L$), with $t_\pm=t_P\pm\frac{1}{\sqrt{4\ln2}}\tau_L[1+\frac{4 \,\Omega(t_P)^2}{e\,\delta^2}].$ As a result, the transfer becomes more adiabatic, with a very low population $P^{(+)}_{adiab}(t)$ transferred to the adiabatic level $\Psi_+(t)$   at $t\sim t_-$. In addition, this population transferred to the upper adiabatic level $\Psi_+(t)$ returns back to the lower adiabatic level $\Psi_-(t)$  at $t \sim t_+$. Almost no population remains in the excited level after  the pulse. 
 
\noindent 
For a sufficiently high $|\delta|$ value, the evolution of the population transfer becomes completely reversible and the population in the excited adiabatic level is such as 
$|\beta(t_P-\delta t)|=|\beta(t_P+\delta t)|$, as  observed in systems $d$ and $f$. The population of the adiabatic excited level can be calculated in the perturbative approximation, leading to $P^{(+)}_{adiab}(t)\sim [\frac{{\dot{\theta}}(t)}{2[E_+(t) - E_-(t)]}]^2=[Q_{adiab}(t)]^2$ (compare rows 2 and 4 in Fig.~\ref{ch5:fig:adiab} for system $d$). For a very large detuning, $|\delta| \gg \Omega(t_P)$, the population of the adiabatic excited level is maximum at $t=t_\pm$ with 
\begin{eqnarray}
 P^{(+)}_{adiab}(t_\pm)=\frac{4\ln2}{e} \left[ \frac{\Omega(t_P)}{\delta^2 \tau_L}\right]^2=\frac{1}{4\,e\ln2}\left[\frac{\Omega(t_P)}{\delta}\frac{\delta\omega}{\delta}\right]^2 
\label{ch5:eq:pop-max-adiab}
\end{eqnarray}
For off-resonant excitation, $|\delta|>\delta\omega$, the maximum   population in the adiabatic level is very small ($P^{(+)}_{adiab}(t_\pm)\ll1$), decreasing with $|\delta|$ more rapidly than the maximum population in the diabatic levels,  equal, in the perturbative limit,  to:  
\begin{equation} 
P_{diab}(t_P)=\left[\frac{\Omega(t_P)}{\delta}\right]^2 .
\label{ch5:eq:pop-max-diab}
\end{equation}
\noindent
To summarize, the three parameters $\delta$, $\delta\omega$ and $\Omega(t_P)$ characterizing the excitation of a two-level system by a Gaussian pulse fully determine the  dynamics. The nearly-resonant or off-resonant character of the process depends on the ratio $\delta/\delta \omega$. For a nearly-resonant excitation of a level $|v'\rangle$ lying within the pulse bandwidth ($|\delta_{v',v_0"}|<\delta \omega$) and in the weak field limit $\Omega(t_P)\ll\delta \omega$, the population transferred to this level  is  $ |b_{v'}(t\rightarrow \infty)|^2= 8\pi \ln{2} \left[\frac{\Omega(t_P)}{\delta \omega}\right]^2 \left[\frac{{\overline{\cal{E}}}(\delta_{v'})}{{\overline{\cal{E}}}(\delta=0)}\right]^2$. Nearly-resonant excitation acquires a strong-field character as soon as $\Omega(t_P)\gg\delta \omega$; then a highly non-adiabatic transfer occurs  in the wings of the pulse, at $t_{na\pm}$. The final population transfer strongly depends on the value of the phase difference accumulated in the adiabatic wavefunctions $\Psi^P_{\pm}(t)$ during the time interval $[t_{na-}, \,t_{na+}]$. This phase difference is  at the origin of the Rabi oscillations observed in the population of the diabatic levels $|g\rangle$ and $|e\rangle$.  

\noindent
For off-resonant excitation, the evolution can be described in the perturbative limit if $\Omega(t_P)\ll \delta$; it results in a completely adiabatic dynamics, with no final population transfer, the population of the diabatic levels following the variation $[f(t)]^2$ of the pulse intensity.

\noindent
Let us emphasize that the conclusions reported above are valid for a  pulse with  sufficiently slow  time-dependences  in both the electric field envelop and the instantaneous frequency. In particular, they are not valid for a spectrally cut Gaussian pulse \cite{albert2008}. In this case, rapid  variations of the instantaneous frequency around the pulse maximum are responsible for a nonadiabatic character of the off-resonant excitation and the off-resonant levels remain  populated after the pulse \cite{merli2009}. 
\subsubsection{\label{ch5:subsubsec:WP-Adiabtic} Excitation of molecular wavepackets}

\noindent
Some information on the adiabaticity of the population transfer in the optimal multi-level system can be obtained by considering directly the excitation of molecular wavepackets in the lower and excited electronic states and reducing the WP description to a two-level problem. Indeed, as we are considering pulses with duration ($\sim100$ fs) much larger than the vibrational period of the considered levels, it is possible to ignore the vibrational motion during the laser excitation, that is to study the excitation process in the impulsive approximation \cite{banin1994}.

\noindent
In the time-dependent Hamiltonian describing the laser excitation of a diatomic molecule in the WP approach, given in Eq.~(\ref{ch5:eq:eqcpl-inst-RWA}),
\begin{eqnarray}
 \hat{H} & = & \left[ \begin{array}{lc}
  {\hat T} + \overline{V}_g(R)+\frac{\hbar \omega_L}{2}& 
  \frac{1}{2}W_{ge}(t) \\
  \frac{1}{2}W_{eg}(t) & 
  {\hat T} +  \overline{V}_e(R)-\frac{\hbar \omega_L}{2}
  \end{array} \right],  
 \label{ch5:eq:hamilto-diabatic}
\end{eqnarray}
\begin{figure}
\centering
\includegraphics[viewport=0 0 360 216, width=7cm, clip]{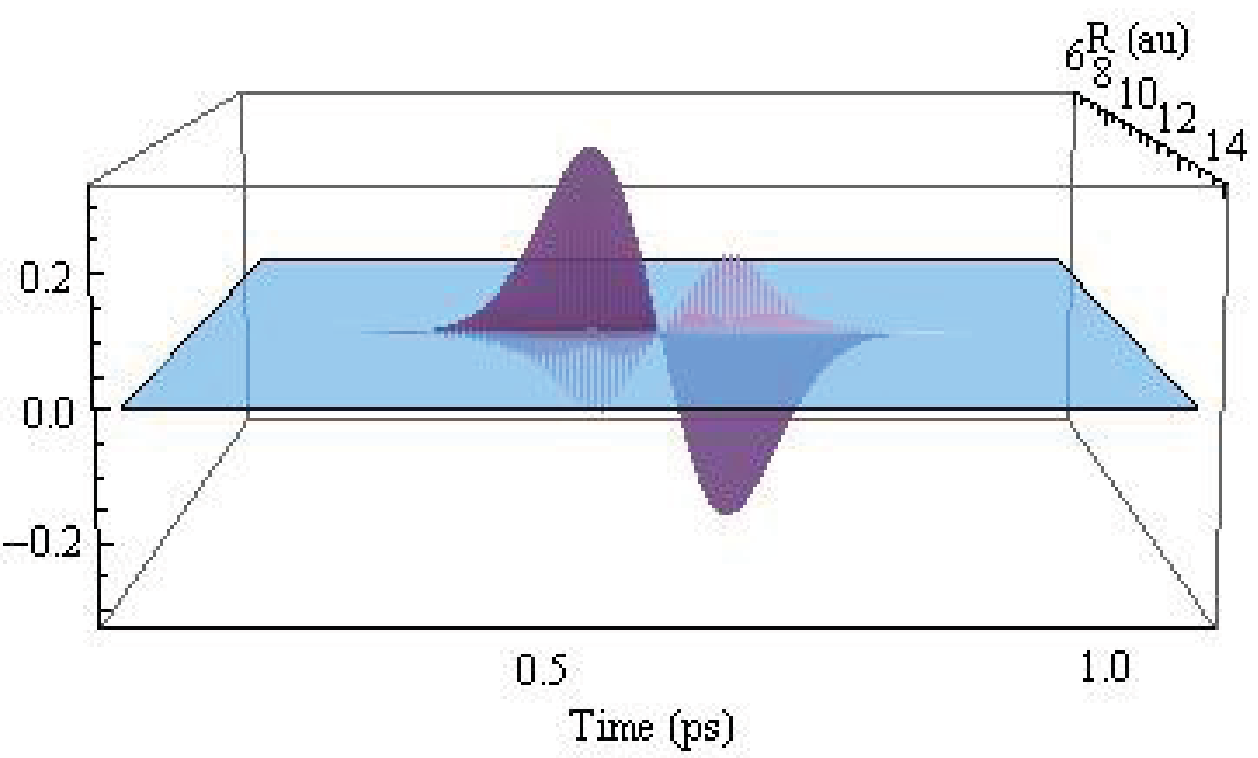}
\includegraphics[viewport=0 0 360 216, width=7cm, clip]{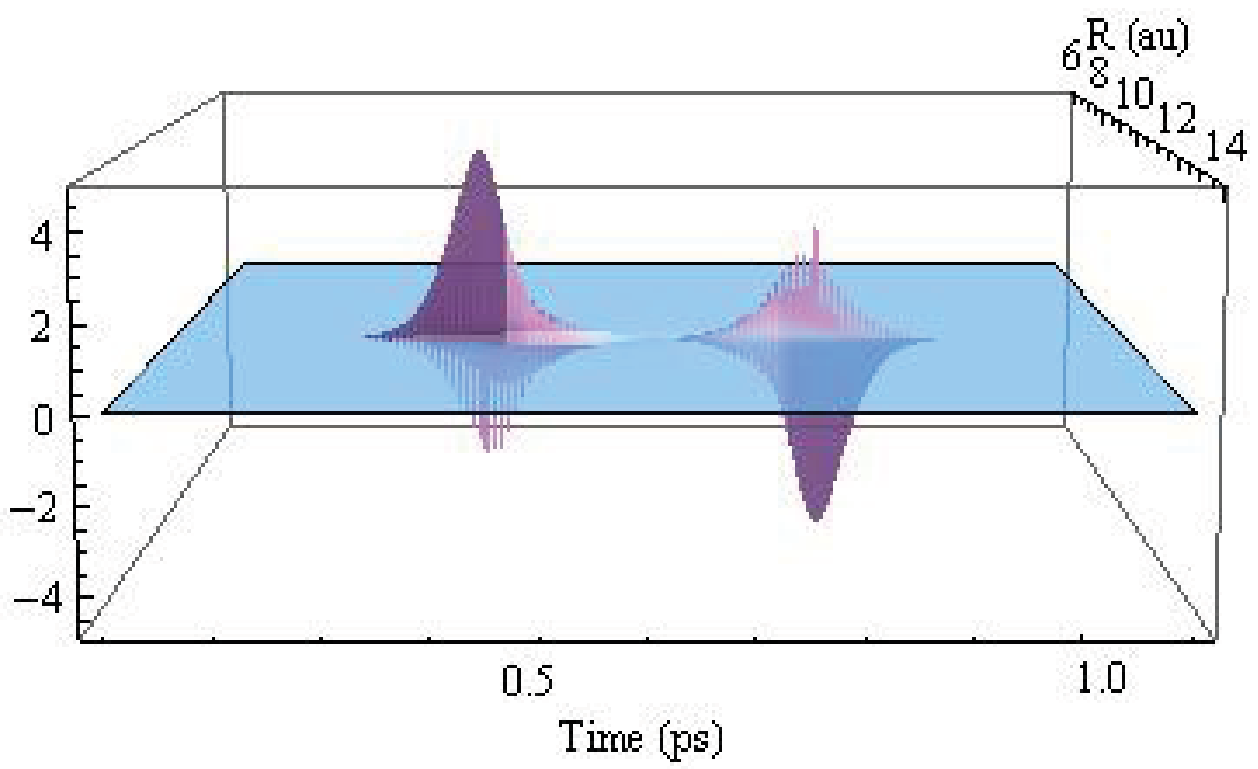}
\caption{\label{ch5:fig:2Levels-adiabCond}(Color online) Criteria of adiabaticity  for the excitation of a molecular wavepacket by a gaussian pulse: variation of  $\overline{Q}_{adiab}(R,t)$ (Sec. \ref{ch5:subsubsec:WP-Adiabtic}) as a function of the internuclear distance $R$ (in au) and of the time $t$ (in ps). The condition for an adiabatic evolution is broken near $R_c=10.5$ au and for times corresponding to the beginning and the end of the Gaussian pulse. Non-adiabatic effects are more important for low-intensity pulse  $W_L^\pi/120$ (upper panel) than for high-intensity pulse  $W_L^\pi$ (lower panel).}
\end{figure}
  
\noindent
we neglect the kinetic energy $\hat{T}$. Introducing the difference between the two dressed potentials $$ \Delta_L(R)=\overline{V}_e(R) -\overline{V}_g(R) -\hbar \omega_L$$ and ignoring the mean potential $$2\underline{V}(R)= \overline{V}_e(R) + \overline{V}_g(R)\; ,$$ which introduces only an $R$-dependent phase factor, we can consider, at each internuclear distance $R$, the two-level Hamiltonian in the diabatic representation \cite{luc2003},
\begin{equation}
 \left[\begin{array}{lc}
 -\frac{1}{2}\Delta_L(R)& 
 \frac{1}{2}W_{ge}(t) \\
 \frac{1}{2}W_{eg}(t) & 
 +\frac{1}{2}\Delta_L(R)
 \end{array} \right] ,
 \label{ch5:eq:definition-DiabR}
\end{equation}
\noindent
and analyze the adiabaticity of the excitation process, by calculating the $R-$ and $t$-dependent function $\overline{Q}_{adiab}(R,t)$ similar to the function ${Q}_{adiab}(t)$ defined in Eq.~(\ref{ch5:eq:Q-adiab}).

\noindent
Figure \ref{ch5:fig:2Levels-adiabCond} shows $\overline{Q}_{adiab}(R,t)$ 
as a function of  $R$ and  $t$. For a pulse with a carrier frequency $\omega_L$ resonant with the transition a$^3\Sigma^+v_0"=37$ $\rightarrow$ $b^3\Pi v_0'=43$, the dressed potentials  cross each other ($\Delta_L(R_c)=0$) at $R_c=10.53$ au. The adiabatic condition is broken at the beginning and at the end of the Gaussian pulse $t=t_P-\tau_L=0.48$ ps and 
$t=t_P+\tau_L=0.72$ ps (small laser intensity). It is also broken at internuclear distance close to $R_c$ (small detuning). This determines both the  times  where population can be transferred from the ground electronic state to the excited electronic state and the spatial location of the transferred population. For the studied unchirped pulse, population transfer occurs around $R_c$ and in the wings of the pulse.
\subsection{\label{ch5:subsec:LbyL-degenerate} Influence on the dynamics of the far-from-resonance levels }
\begin{figure}
\centering
\includegraphics[viewport=0 0 556 608, width=8cm, clip]{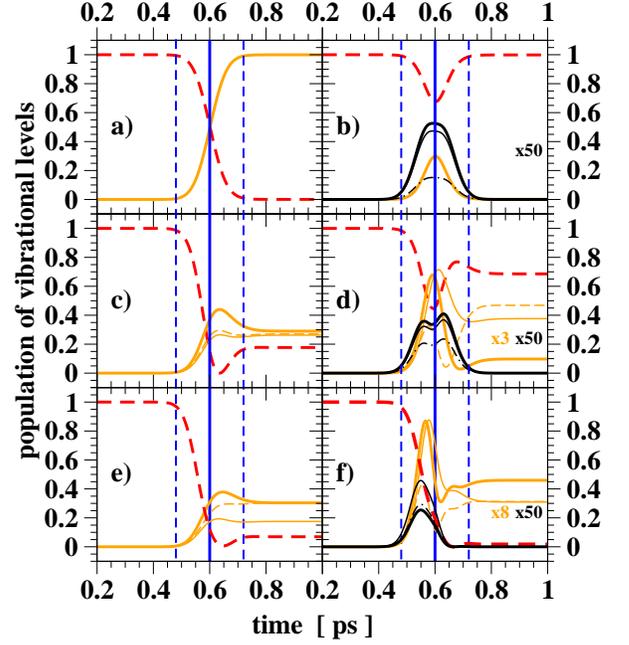}
\caption{\label{ch5:fig:increasing-basis}(Color online) LbyL calculation: influence on the dynamics of the far-from-resonance $b^3\Pi v'$ levels, when only the level $v_0"=37$  is introduced in the lower state. Variation with time (in ps) of the populations of the vibrational levels $b^3\Pi$ $v'$ with $v'=43$ (thick continuous orange [gray] line), $v'=42$ (thin continuous orange [gray] line), $v'=44$ (thin dashed orange [gray] line), $v'=195$ (thick black line), $v'=194$ (thin black line), $v'=196$ (thin dot-dashed black line) and of the initially populated vibrational level $a^3\Sigma^+$ $v_0"=37$ (dashed red [black] line). Some populations have been multiplyied by the factor indicated in the figure. The pulse maximum at $t_P$ (vertical continuous blue [black] line) and duration indicated by the vertical  dashed blue [black] lines satisfies the $\pi$-pulse condition.  Different basis sets  $[v",\,v']$ are used.  The  difference between lefthand and righthand panels of the same line is the absence or presence of far-from-resonance levels. Upper line, a) $[v_0"=37, \, v_0'=43]$ and b) $[v_0"=37, v_0'=43$ and $v'=194-196]$; middle line, c) $[v_0"=37, \, v'=42-44]$ and d) $[v_0"=37, \, v'=42-44$ and $v'=194-196]$; lower line, e) $[v_0"=37, \, v'=13-60]$ and   f) $[v_0"=37, \, v'=0-218]$.}
\end{figure}
We pay now a particular attention to the contribution to the dynamics coming from far-from-resonance levels. We first analyze the dynamics of excitation in a multi-level system including both nearly-resonant and far-from-resonance excited levels $bv'$ while keeping only the single level $av_0"$ in the lower electronic state. Then  we  introduce all the lower  $[v"=0-204]$ levels of the optimal set to obtain a complete view of the modifications of the dynamics of close to the resonance levels induced by  far-from-resonance levels. 

\begin{figure}
\centering
\includegraphics[viewport=0 0 600 500, width=8cm, clip]{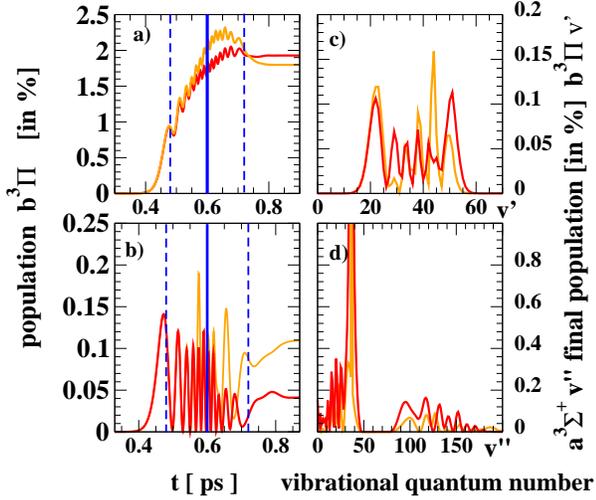}
\caption{\label{ch5:fig:bfar-modif}(Color online) LbyL calculations: influence on the dynamics of  far-from-resonance $b^3\Pi$ $v'$ excited levels, when all the levels $a^3\Sigma^+$ $v"$ of the set $C$ are introduced in the lower state. Basis sets $[v"=0-204, \,v'=13-60]$ (orange [gray] lines) and $[v"=0-204, \, v'=0-218]$ (red [black] lines). The  pulse (vertical continuous and dashed blue [dark] lines) satisfies the $\pi$-pulse condition. Left column: variation with time (in ps) of the population of the  $b^3\Pi$ $v'$ levels (in percentage); panel a): total population in the levels $v'=13-60$; panel b) population in the resonant level $v_0'=43$. Right column: final distribution of population (in percentage) in the excited levels $v'$  (panel c)) and lower levels $v"$ (panel d)) with a maximum, for $v_0"$, equal  to 90.9 and 76.0 for the first and second basis sets respectively.}
\end{figure}
\subsubsection{\label{ch5:subsec:LbyL-ex} A single level in the lower electronic state}

\noindent
Starting with a single $v_0"=37$ level in the ground electronic state, we progressively grow  the basis set in the excited state, by adding either nearly-resonant levels, {\it{i.e.}} close to the oblique transition, or far-from-resonance levels, {\it{i.e.}} close to the vertical transition. 

\noindent
In the upper row (Figs.~\ref{ch5:fig:increasing-basis}a,b), we compare the two-level system that consists only of the two resonant levels $[v_0"=37, v_0'=43]$ to the small 5-level system $[v_0"=37, v'=43,194-196]$ containing the two resonant levels and the three far-from-resonance levels $v'=194-196$. As expected, for the resonant two-level system excited by a $\pi$-pulse, a total exchange of population is observed. For the  basis set b), a very low population is transferred during the pulse to the three additional levels, less than 1\%. Nevertheless the presence of these levels modifies completely the dynamics of the population in the $v_0'=43$ resonant level: only 29\% of the population is transferred in this level at $t=t_P$, instead of 50\% in the two-level system, and, in addition, the population disappears almost completely (0.012\%) at the end of the pulse. In fact, for the two  $v_0"$ and $v'_0$ levels, which are degenerate in the diabatic representation (Eq.~(\ref{ch5:eq:h(t)}) with $\Delta^e_{v'=43,v_0'}=0$ Eq.~(\ref{ch5:eq:Delta-g-e})), the maximum coupling strength, $\Omega_{v_0'=43,v_0"}= 2.10\times 10^{-4}$ au, is of the same order of magnitude as the second order contribution $[\Omega_{v'=195,v_0"}]^2/\Delta^e_{v'=195,v_0'}=3.4\times 10^{-4}$ au corresponding to the vertical transition $v'=195$. This modifies significantly the energies of the instantaneous adiabatic levels connected to the resonant $v_0"=37$ and $v_0'=43$ diabatic levels and therefore the phase difference accumulated between $t_{na-}$ and $t_{na+}$ in the adiabatic wavefunctions $\Psi^P_{\pm}$ (Eq.~(\ref{ch5:eq:psi-adiab-pulse})) or, during the pulse, the beating between the probability amplitudes of  the resonant levels  (Sec. \ref{ch5:subsubsec:adiab-resonant}). This explains qualitatively the strong changes in the dynamics of the excitation process.

\noindent
In the middle row  (Figs.~\ref{ch5:fig:increasing-basis}c,d), the two nearly-resonant levels $v'=42$ and $v'=44$ are added to each above described basis set. For these levels, the overlap integrals with the $v_0"$ level (0.029 and 0.031) are nearly equal to the overlap integral (0.034) of the resonant transition, and the detunings are small $\pm0.0002$ au. In the system $[v_0", \, v'=42-44]$ introducing only the nearly-resonant levels, the population transferred to the excited state is shared between the three excited levels with a large total transfer (82\%). When the three levels close to the vertical transition $v'=194-196$ are added, there is a low transfer (3.3\%) to the resonant $v_0'=43$ level   but a larger transfer (12.5 and 15.6\%) to the nearly-resonant levels $v'=42$ and $v'=44$. Here also the introduction of the far-from-resonance levels modifies the excitation dynamics of the nearly-resonant levels. In particular, there is a strong decrease in the population transferred to the excited electronic state at the end of the pulse, 31.4\% instead of 82.3\%, and an important change in the branching ratios in the population of the nearly-resonant levels. 

\noindent
In the lower row  (Figs.~\ref{ch5:fig:increasing-basis}e,f), larger basis sets are introduced in the excited state. The set e) $[v_0"=37, \,v'=13-60]$ consists on all the excited levels remaining populated after the pulse (Fig.~\ref{ch5:fig:3-groups}) in the WP calculation. The dynamics of excitation of the nearly-resonant levels $v'=42-44$ is qualitatively the same as in set c), with a total population transfer equal to 78.2\%, but with a change in the branching ratios. This shows that, in this group of  $v'=13-60$ levels, the nonadiabatic dynamics (Sec. \ref{ch5:subsubsec:2-types}) is dominated by the three nearly-resonant levels. For the set  f) $[v_0"=37, \,v'=0-218]$ introducing a still larger basis in the excited state, the dynamics of the excitation of the nearly-resonant levels  differs from that one observed for the set d) $[v_0"=37, \,v'=42-44,194-196]$. The final population in the nearly-resonant levels $v'=42-44$ decreases from the value 31.4\% to the value  13.5\%, showing that far-from-resonance levels other than the $v'=194-196$ ones contribute to the dynamics. For the two basis sets e) and f), 
 almost all the  population is transferred to the excited electronic state, 93.1\%  and 98.24\% respectively (see the low value of the final population of the initial level $v_0"=37$),  this population being mainly distributed in  the nearly-resonant levels $v'\sim 42-44$ for  set e), but in lower levels $v'\sim 36-40$ for set f). 
\subsubsection{\label{ch5:subsubsec:analysis-level-2} Several levels in the lower electronic state}

\noindent
In this section we incorporate all the levels $v"=0-204$ of the lower electronic state included in the optimal basis set (Sec. \ref{ch5:subsubsec:large-basis}) and we analyze how the far-from-resonance levels modify the excitation dynamics. We consider the basis sets $[v"=0-204, \,v'=13-60]$ and $[v"=0-204, \, v'=0-218]$. The first set encompasses  only the excited levels which remain populated after the pulse in the WP treatment whereas the second one is the optimal set. In Fig.~\ref{ch5:fig:bfar-modif} we show the computed time-evolution of the total populations of the levels $v'=13-60$ (Fig.~\ref{ch5:fig:bfar-modif}a) and of the resonant level $v_0'=43$ (Fig.~\ref{ch5:fig:bfar-modif}b). We display the final distributions of population in the vibrational levels of the excited (Fig.~\ref{ch5:fig:bfar-modif}c) and the ground  (Fig.~\ref{ch5:fig:bfar-modif}d)  electronic states.
 
At the beginning of the pulse,  for $t \le 0.53 $ ps, when the pulse intensity increases, weak population recycling has yet occured and the contribution to the dynamics of far-from-resonance levels is not very important. When the pulse reaches its maximum intensity at $t=t_P=0.6$ ps, the population exchange between the lower and excited electronic states becomes more important and noticeable changes become observable in the Rabi oscillations occurring either in the population of the resonant level $v_0'=43$ or in the total population of the levels $v'=13-60$. When far-from-resonance levels are introduced in the basis set, the total population transferred during the pulse to the levels $v'=13-60$ is smaller; nevertheless there is no significant change in the total population transferred to the excited state (1.9\% instead of 1.8 \%)(Fig.~\ref{ch5:fig:bfar-modif}a). Concerning  the Rabi oscillations of  the population of the resonant level $v_0'=43$ (Fig.~\ref{ch5:fig:bfar-modif}b), some modifications occur for $t>t_P$ and half of the population remaining in this level is transferred back to the ground state (0.04\% instead of 0.11\%). We remind the reader that a similar decrease in the population of the resonant level induced by including far-from-resonance levels has already been observed in the simple LbyL models discussed  in Fig. \ref{ch5:fig:increasing-basis}). Concerning the final distribution  in the excited state population(Fig.~\ref{ch5:fig:bfar-modif}c), the population is, on average, shifted toward slightly higher $v'$-values. In the ground state (Fig.~\ref{ch5:fig:bfar-modif}d), the population is spread over a larger energy domain, with a smaller population transferred back to the initial level $v_0"=37$ (76.0\% instead of 90.9\%).    
\subsection{\label{ch5:subsubsec:analysis-quasi-degenerate}Blockade of the excitation due to a quasi-degenerate level in the lower electronic state}

\noindent
When the basis set $[v_0"=37,\,v'=0-218]$ is used (Figs.~\ref{ch5:fig:increasing-basis}e,f), the  resonant level $v_0'=43$ population does not exhibit  a large number of Rabi oscillations contrary to what is observed in Fig.~\ref{ch5:fig:a-b-levels-d} using the optimal set. Furthermore there is a strong transfer of population to the excited electronic state (98.2\%), substantially  different from the weak transfer (1.9\%) obtained with the optimal set. In this section, we  analyze in more detail the specific role of the levels of the ground electronic state, especially those which are quasi-degenerate with the initially populated one. In the energy range close to this initial level, the spacing between consecutive bound levels is indeed much smaller than the laser bandwidth $\delta \omega=120$ cm$^{-1}$. All bound levels with $v"\ge 34$ and all continuum states with an energy up to 115 cm$^{-1}$ (with the chosen grid, discretized scattering levels up to $v"=134$) are such that $|\Delta^g_{v",v_0"}| < \delta\omega$.  

\noindent
To analyze the excitation dynamics from  a group of quasi-degenerate levels, we consider in  Appendix \ref{ch5:subsec:N-levels} a simple model describing the excitation of a single sublevel from an $N$-fold degenerate level, which admits an analytical solution. The comparison with a non-degenerate two-level system shows that, in the high field regime, the population transfer is divided by ${N}$, whereas the remaining population is equally distributed among the $N$ sublevels of the degenerate lower level. When the number $N$ increases, the transfer of population toward the excited level decreases: there is a blockade of the excitation induced by the degeneracy of the lower level, with no transfer at all for $N \rightarrow\infty$.

\begin{figure}
\centering
\includegraphics[viewport=0 0 610 570, width=8cm, clip]{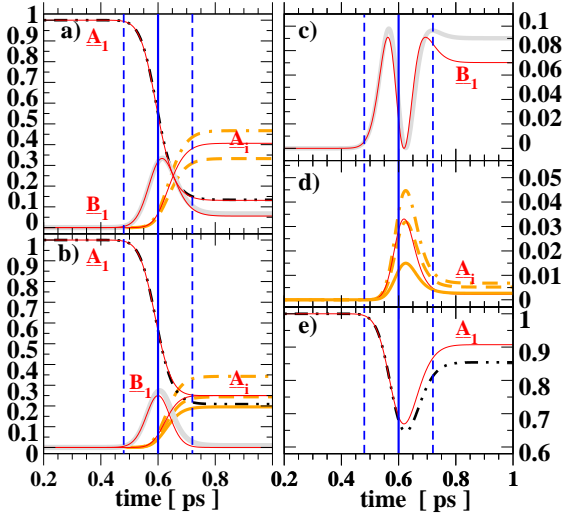}
\caption{\label{ch5:fig:11levels}(Color online) Influence of a quasi-degenerate level in the ground electronic state. Evolution of the populations as a function of time (in ps), under a pulse (continuous and dashed blue vertical lines) satisfying the $\pi$-pulse condition $W=W_L^{\pi}$. The results obtained using the LbyL method are compared to the analytical formulas (Eq.~(\ref{ch5:eq:a-Deg1})) describing the excitation from a $N$-fold degenerate level (thin continuous  red [black] lines), where the initial population is in a single level ($\underline{A}_1$) of the manifold, the other $N-1$ sub-levels  ($\underline{A}_i$) being unpopulated; the excited level ($\underline{B}_1$) is resonantly excited. Panel a) $N=3$: basis set $[v"=36-38,\,v_0'=43]$. Panel b) $N=4$: basis set $[v"=36-39,\,v_0'=43]$. Panels c), d), e) $N=11$: basis set $[v"=32-42,\,v'=43]$. Initially populated level $a^3\Sigma^+ v_0"=37$ (double-dot dashed black line) and resonantly excited level $b^3\Pi v_0'=43$ (thick continuous clear-gray  line]. Non resonant $a^3\Sigma^+$ levels described by the same formula in the analytical model:  $v"=36$ (medium-thick dot-dashed orange [gray] line); $v"=38$ (medium-thick dashed orange [gray] line); $v"=39$ (medium-thick  continuous orange [gray] line).}
\end{figure}

\noindent
As the first example we consider two basis sets, $[v"=36-38,\, v'=43]$ and $[v"=36-39,\,v'=43]$, and the $\pi$-pulse resonant with the transition $v"_0=37 \rightarrow v'_0=43$ (Figs.~\ref{ch5:fig:11levels}a,b). For the set a) where  $N$=3, the final population transferred to the $v_0'$ level is equal to 0.0556. The populations of the initial $v_0"$ level and of the other two ground levels are respectively equal to 0.1313 and 0.4065. For the set b) with $N=4$, no population remains after the pulse in the excited level $v_0'$, the effective pulse area being equal to $\frac{\sqrt{N}}{2}\pi=\pi$. Simultaneously, the four ground state levels are equally populated (population  $\frac{1}{4}$). At the maximum of the pulse, when $\frac{\sqrt{N}}{2}\Theta(t_P)=\frac{\pi}{2}$, the population in the excited level is equal to $\frac{1}{N}=\frac{1}{4}$. In the initially populated level $v_0"$ it is equal to $\frac{9}{16}$ and in the other three ground levels to $\frac{1}{16}$. 
The evolutions of the populations in the quasi-degenerate case agree almost perfectly with the $N$-fold degenerate model with $N=3$ and $N=4$. 

\noindent
More realistic results, obtained by using the LbyL approach with the basis set $[v"=32-42,\,v_0'=43]$, are also presented  (Figs.~\ref{ch5:fig:11levels}c,d,e), together with comparable results of the analytical model. For the ground levels $|\Delta^g_{v",v_0"}|\le0.000027$ au and the overlap integrals vary in the range 0.018-0.040. For this system, which includes a $11$-fold quasi-degenerate lower level, the dynamics of excitation is similar to that of a degenerate level with $N=11$. The effective area for the pulse is $\sqrt{N}\pi=3.3\pi$, and, during the pulse, the population of the resonantly excited level oscillates between 0 and $\frac{1}{N}=0.09$. Simultaneously, the population is redistributed among the quasi-degenerate levels of the ground electronic state. 

\noindent
To summarize, in the strong-field excitation from a level close to the dissociation threshold, the high density of levels in the initial state is at the origin of a blockade of the excitation process. Simultaneously the increase of the effective Rabi frequency explains the oscillations occuring during the pulse in the population of the quasi-resonant levels. Let us remark that this phenomenon is similar to the ionization suppression occuring in the Rydberg atom ionization by an intense laser pulse.  When  the $N$  initial discrete levels are exactly degenerate, only $1/N$ of the initial population ionizes in a time  divided  by the  factor $N$ \cite{parker1990}.

\noindent
Finally we remark on the influence of the far-from-resonance excited levels on the excitation from quasi-degenerate lower levels. As expected, the basis set $[v"=0-204, \,v_0'=43]$ yields a blockade of the excitation: the final population of the initial level $v_0"=37$ is large, amounting to 90.6\% and, simultaneously, a very weak population, equal to 1.0\%, is transferred to the resonant level $v_0'=43$.  
When adding some far-from-resonance excited levels, in the basis set $[v"=0-204, \,v'=43,194-196]$, one observes simultaneously a blockade of the excitation and an important redistribution of population within the ground state:  the population transferred to the resonant level $v_0'=43$ is almost negligible (1\%), but 78.5\% of the population is  redistributed among the levels $20\le v"\le 47$ and 20.5\% in the continuum, mainly in scattering levels with an energy smaller than 20 cm$^{-1}$. Here also, the contribution of the far-from-resonance levels $v'=194-196$ is crucial. These levels are only weakly populated during the pulse, but their population is recycled back to a large number of vibrational levels of the ground electronic state.
\section{\label{ch5:sec:ImproveT}Discussion and perspective; train of pulses}

\noindent
In this paper, we have explored the possibility of enhancing the rate of formation of stable RbCs molecules in the absolute ground level Rb(5s)Cs(6s) $X^1\Sigma^+v" = 0$. More precisely, we have analyzed the excitation  by a single unchirped Gaussian pulse of molecules already formed in weakly bound levels of the Rb(5s)Cs(6s) $a^3\Sigma^+$ state, after photoassociation of ultracold Rb and Cs atoms followed by spontaneous radiative decay. When the final level of the excitation is a spin-mixed level, for example a level of symmetry $0^+$ or $1$, it will be possible to transfer optically, in a second step, the population from this electronic excited level to the absolute ground level. Restricting the description to uncoupled electronic states in the Hund's case $a$ coupling scheme, we have investigated the possibilities offered by the presently widely developed femtosecond laser sources, to transfer efficiently the population from the excited $a^3\Sigma^+ v_0" = 37$ level toward the $b^3\Pi v_0' =43$ level. 
 
\noindent
The dynamics of the photoexcitation process is modeled with the WP method, calculating the time-evolution of the wavepackets propagating along the  electronic states coupled by the laser pulse. This employed non-perturbative method allows us to analyze dynamics all the way from the low- to the high-intensity regime, the latter being the $\pi$-pulse for the resonantly-driven  $v_0" = 37$ $\rightarrow$ $v_0' =43$ transition. We have also developed a LbyL description, and employed a variety of subsets of vibrational levels in numerical simulations. The comparison between various restricted LbyL calculations and with the full-scale WP results allowed us to qualitatively understand the complex evolution of the population of the multitude of  vibrational levels and to identify the specific influence  of these levels on the dynamics. 

\noindent
In the perturbative limit, only the quasi-resonant levels lying within the bandwidth of the intensity distribution are excited. For a not too short duration of the pulse $\tau_L$ corresponding to a bandwidth $\delta\omega$ smaller than the energy spacing of the vibrational levels in the excited electronic state, the excitation process is selective. Conversely, the efficiency of the population transfer is very low.  

\noindent
With increasing intensity, more levels are excited, but the excitation rates remain very small and the dynamics becomes more adiabatic. We have shown that, in the strong-field regime, {\it i.e.} for the $\pi$-pulse, far-from-resonance excited levels with high Franck-Condon factors are populated during the pulse. At the maximum of the pulse intensity, the population of the levels excited through a vertical transition at the outer turning points of their wave functions is much larger than the population of the quasi-resonant levels. Because of the adiabatic character of the excitation of far-from-resonance levels, the time-dependence of their population follows the smooth evolution of the pulse intensity, notwithstanding the high value of the Rabi frequency, and after the pulse  no population is retained in these levels. Nevertheless these levels contribute to the excitation process, in particular they give rise to a large population recycling and to an important  population redistribution in the ground electronic state. The individual contributions to the amplitude of population of a particular level arising from the other levels are very intricate, making  the analysis  and therefore the control of the excitation process with unchirped pulses in the high-field regime very difficult. It is worth noticing that quantum control of molecular wavepackets could still be possible in the strong-field regime \cite{araujo1999}, but in very specific conditions: for 'vertical' transitions with high Franck-Condon factors, not involving quasi-degenerate lower levels, and imposing special shape requirements to the standing edge of the pulse \cite{dubrovskii1992}, in order to avoid population redistribution in the lower state.

\noindent
Furthermore, for the system under study, the initial level $a v_0"=37$ lies very close to the dissociation threshold, where the density of levels is very high. We have shown that this situation is at the origin, in the high-field regime, of the excitation blockade. A simple model describing the resonant excitation of a single level from an $N$-fold degenerate level is developed, which reproduces  comparable LbyL calculations well. In the strong-field regime, the  excited level population is governed by a pulse area larger by a factor of $\sqrt{N}$ than the real area of the pulse and its amplitude is divided by $N$. This explains the oscillations with a low amplitude observed, during the pulse, in the population of the resonantly-excited  level. Due to this blockade phenomenon, high intensity  and large bandwidth pulses  are poorly suited for gaining high excitation rates. 
  
\noindent
In order to increase the population transfer to the resonant level $v_0'=43$ while conserving the selectivity provided by the low field regime, we have explored the possibility of using a weak intensity train of ultrashort coherent pulses. We recapitulate main features of  such trains  in the time and frequency domains  in Appendix \ref{app:pulse-train}. Previously,
 coherent excitation of a two-level system by a train of short pulses has been described analytically \cite{vitanov1995}. Transient coherent accumulation for two-photon absorption via an intermediate level has been demonstrated in atomic Rb \cite{stowe2006}. Application to  efficient selective vibrational population transfer between electronic states of a diatomic molecule has been discussed by Araujo \cite{araujo2008}. 
\begin{figure}
\centering
\includegraphics[viewport=0 0 610 573, width=8cm, clip]{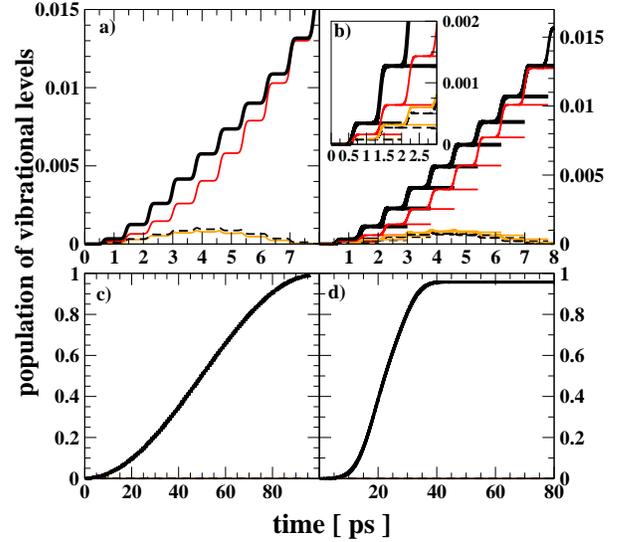}
\caption{\label{ch5:fig:accumulative-p}(Color online) Panels a), b), c): Excitation of the resonant transition $a^3\Sigma^+ \; v_0"=37 \rightarrow b^3\Pi v_0'=43$ in the low field regime by successive Gaussian femtosecond pulses  with duration $\tau_L=0.12$ ps, repetition time $T_{rep}=0.8$ ps and  maximum coupling strength $W_L^\pi/120$. Variation with time (in ps) of the population of the excited electronic state $b^3\Pi$ (thick black line), of the resonant level $b^3\Pi v_0'=43$ (thin red [black] line)  and of the quasi-resonant levels $v'=42$ (thin orange [gray] line) and $v'=44$ (thin black dashed line) levels. Panel a): train of $\mathcal{N}=10$ coherent pulses in the LbyL approach with the basis set $[v_0"=37,\, v'=42-44]$. Panel b): train of $\mathcal{N}=10$ pulses  in the WP approach, where the lower  and  excited wavepackets obtained in the middle of the interval between the pulses $P_i$ and $P_{i+1}$ are taken as  initial wavepackets for the pulse $P_{i+1}$. Panel c):  same as for  panel a), but for a train of $\mathcal{N}=120$ pulses, corresponding to an effective $\pi$-pulse.  Panel d): single picosecond $\pi$-pulse pulse with duration $120\tau_L=14.4$ ps and maximum coupling strength $W_L^\pi/120$ in the LbyL approach with the same basis set.}
\end{figure}

\noindent
In a preliminary study, we  analyzed the dynamics of excitation by a train of pulses in the perturbative regime. Each individual pulse has the Gaussian shape of duration $\tau_L=0.12$ ps and of maximum coupling strength $W_L^\pi/120$. The repetition time is $T_{rep}=0.8$ ps, slightly smaller than the vibrational period $T^{vib}_{e,v_0'}$ in the excited state (Sec. \ref{ch5:subsec:pi-pulse-cond}) and with a vanishing pulse-to-pulse carrier-envelope-offset phase shift $\Delta\phi_{ce}$. We  performed calculations both in the $WP$ and the LbyL approaches. In the WP calculations, the final state at the end of each individual pulse is taken as the initial condition for the following pulse. The total population transferred to the excited electronic state $b^3\Pi$ and its distribution among the different $v'$ levels during  $\mathcal N=15$ pulses are shown in Fig.~\ref{ch5:fig:accumulative-p} (upper right panel). We also carried out LbyL calculations using the basis set $[v_0"=37, \, v'=42-44]$.
The computed time-variation of the population in the excited levels  is presented in Fig.~\ref{ch5:fig:accumulative-p} (left panels). The LbyL calculations reproduce perfectly the WP results, the small LbyL basis set clearly being sufficient for  low-intensity pulses.
  
\noindent
Pulse after pulse, there is an accumulation of the total population transferred to the excited electronic state $b^3\Pi$. When the number of pulses, $\mathcal N$, increases, the distribution of population among  the excited levels $bv'$ becomes more selective, with an accumulation of population in the resonant level $v_0'$. A given pulse transfers to the level $bv'$ an probability amplitude which interferes with the already present probability amplitude, transferred by the previous pulses. The nature of the interferences depends on the phase $\exp[i T_{rep} \delta_{v',v_0"}]$ involving the detuning of the considered  $bv'$ level \cite{araujo2008}. For the resonant level $bv_0'$ the interferences are constructive and the population increases with the growing number of pulses. For other vibrational levels, due to the mismatch in this phase, the population will oscillate with $\mathcal N$, without experiencing accumulation. This increase of the selectivity of the excitation with the number of pulses is a signature of  the comblike structure of the energy spectrum ${\overline{\cal{E}}}(\omega)$ (Eq. (\ref{ch5:eq:train-P-freq})) of the pulse train. The frequency  spectrum consists of equally spaced ``teeth'', with a spacing proportional to the repetition frequency $T_{rep}^{-1}$, an intensity proportional to $\mathcal N^2$ and a width narrowing as $\mathcal N^{-1}$.  

\noindent
When the number of pulses becomes equal to $120$, the total population initially in the $v_0"=37$ level is transferred to the resonant excited level $v_0'=43$. Let us remark that the total pulse area for the train with $120$ pulses, each with a pulse area equal to $\pi/120$, amounts to $\pi$. This train of pulses is thus equivalent to a single $\pi$ pulse in the perturbative regime. If the number of pulses continues to increase, the population cycles back to the initial level. The total duration of the pulse train amounts to $96$ ps, much smaller than the lifetime of the resonant level, which is smaller than 30 ns \cite{londono2011}. 

\noindent
In conclusion, to succeed in controlling the dynamics of  photoexcitation with unchirped femtosecond lasers, it seems necessary to employ  
low-intensity pulses. The radiative lifetime of the excited level can be disregarded if the total duration $T_{tot}={\cal{N}} T_{rep}$ of the pulse train is sufficiently small. Consequently, the repetition rate has to be as large as possible. We notice the technological developments aimed at increasing the repetition rates in the train of femtosecond or a few picosecond pulses using acousto-optic devices are in progress. Repetition frequencies up to 50 GHz were obtained with electro-optic phase-modulator shaping of a picosecond laser \cite{thomas2010}, much faster than those of the order of 100 MHz obtained from a Kerr lens mode-locked femtosecond  Ti-sapphire laser \cite{stowe2006} or of 100 kHz for a regenerative amplifier seeded by a Mira oscillator \cite{weise2009}. 

\noindent  
For completeness, we mention here that another way to obtain high transfer rate with high selectivity relies on the use of a single pulse in the picosecond domain with a sufficiently narrow bandwidth. To illustrate this point, we computed excitation by a single Gaussian pulse, resonant with the transition $v_0" \rightarrow v_0'$, with maximum coupling $W_L'=W_L^\pi/120$ and duration $\tau_L'=120 \times \tau_L=14.4$ ps (Fig.~\ref{ch5:fig:accumulative-p}, lower right panel). 
Calculations were done in the LbyL method using the basis set $[v_0"=37,\, v'=41-44]$.
The bandwidth of this pulse, equal to $\delta\omega/120=1$ cm$^{-1}$, is sufficiently narrow to include only the single $v_0'$ level within its bandwidth. Its pulse area, proportional to $\tau_L'\times W_L'$ (Eq.~(\ref{ch5:eq:area-pulse})) is equal to $\pi$. Therefore, for this pulse, the population transfer occurs only toward the resonant level and is complete. The corresponding laser sources are not numerous but are presently developed. 

\noindent
Let us emphasize that unchirped femtosecond pulses have been considered throughout this paper. Methods for executing robust, selective and complete transfer of population between a single level and preselected superpositions of levels are presently rapidly developing, both theoretically \cite{shapiro2009} and experimentally \cite{zhdanovich2009}. Such tranfers are obtained through adiabatic passage with intense femtosecond pulses, shaped in amplitude and phase.

\noindent
We finally mention that the possibilities offered by the implementation of a STIRAP (stimulated Raman adiabatic passage) process using femtosecond pulses, instead of the currently used pulses in the microsecond domain, remain to be investigated. Ultimately one would want to produce absolute ground state molecules from weakly-bound molecules formed after photoassociation and spontaneous radiative decay. Keeping this goal in mind, one may want to investigate schemes~\cite{peer2007} where a coherent train of weak pump-dump pairs of shaped femtosecond pulses are used. In that scheme each pair of pump-dump pulses  drives narrow-band Raman transitions between vibrational levels avoiding spontaneous emission losses from the intermediate state.
\appendix
\section{\label{app:MFGH}Complete set of vibrational states from the MFGH method}
The MFGH method is based on the Fourier Grid Hamiltonian method (FGH) with the introduction of an adaptive coordinate, related to the local de Broglie wavelength, to represent the interatomic distance $R$ describing the vibration of the diatomic molecule in the potential $V(R)$. The employed  spatial grid has a  few points $N$ but a large extent $L$ \cite{kokoouline1999}. The Hamiltonian $H_{mol}$ is represented on this grid using a sine expansion  rather than the usual Fourier expansion, in order to avoid the occurrence of ghost levels \cite{willner2004}. For a single channel problem 
 the diagonalization of the Hamiltonian matrix  provides a complete set of $N$ vibrational wave functions $\varphi_v(R)$ ($0\le v\le N-1$) describing bound levels and  discretized continuum states normalized to unit on the grid. As discussed in Ref.~{\cite{londono2011}}, only a small number of scattering wave functions, called 'physical scattering levels', have a realistic behavior throughout the grid. The other ones, which have a high probability density  at short internuclear distance, ensure the completeness of the set for $0\le R \le L$.  The eigenfunctions $\varphi_{v}(R)$ are orthogonal within the box:
\begin{equation}
\int_0^L \varphi_{v_1}(R) \varphi_{v_2}(R) dR=\delta_{v_1,v_2}.
\label{ch5:eq:ortho}
\end{equation}
\noindent
They satisfy the following closure relations, valid for $R\le L$, $R'\le L$:
\begin{eqnarray}
\sum_{v=0}^{{N}-1} \varphi_{v}(R) \times \varphi_{v}(R')&=& \delta(R-R') .
\label{ch5:eq:compl}
\end{eqnarray}
\noindent
In the present paper, the same spatial grid is used for both $a^3\Sigma^+$ and $b^3\Pi$ electronic states. It contains $N=511$ points with of length $L=1258$ $a_0$. The lower and excited electronic states possess 48 and 219 bound vibrational levels. The 'physical scattering levels' describe a very small energy domain, less than 0.01 cm$^{-1}$, above the dissociation limit located at $E=0$. The remaining ones, the 'unphysical scattering levels', cover a large energy range up to 35000 cm$^{-1}$. 
\section{\label{app:dyn}Time-dependent study of photoexcitation in a diatomic molecule}
\subsection{\label{app:WP}Laser-coupled electronic states: 'Wavepacket' description}
\noindent
The evolution of the wavepackets is studied in the rotating wave approximation (RWA) \cite{bookCohen}, by introducing a frame rotating at the angular frequency $\omega_L/2\pi$, which allows one to eliminate rapidly oscillating terms in the system of coupled equations (Eq. \ref{ch5:eq:matricial2x2}). The new radial wave functions corresponding to the lower $\overline{V}_g$ and the excited $\overline{V}_e$ potentials, are defined by:
\begin{eqnarray} 
{\Psi}_g(R,t)&=& \chi_g(R,t)\exp(-i\omega_L t/2) ,\nonumber \\
{\Psi}_e(R,t)&=& \chi_e(R,t)\exp(+i\omega_L t/2) .
\label{ch5:eq:WP-RWA}
\end{eqnarray}
\noindent
Neglecting the high frequency components $\sim\pm 2\omega_L/2\pi$, one obtains the following coupled radial equations:
\begin{eqnarray}
\label{ch5:eq:eqcpl-inst-RWA} 
 &&i\hbar\frac{\partial}{\partial t}\left(\begin{array}{c}
 \Psi_{g}(R,t)\\
 \Psi_{e}(R,t)
 \end{array}\right)\\
 &&=
 \left(\begin{array}{lc}
 {\hat T} + V_g(R)& 
 \frac{1}{2}W_{ge}  \nonumber \\
 \frac{1}{2}W_{eg} & 
 {\hat T} +  V_e(R)    
 \end{array} \right) 
 \left( \begin{array}{c}
 \Psi_{g}(R,t)\\
 \Psi_{e}(R,t) 
 \end{array}\right),
\end{eqnarray}
\noindent
where ${\hat T}$ is the kinetic energy operator and the potentials dressed by the laser frequency $V_g(R)$ and $V_e(R)$,  are given by:
\begin{eqnarray}                                      
 V_g(R) &=& \overline{V}_g(R)+\hbar\omega_L/2,\nonumber \\
 V_e(R) &=& \overline{V}_e(R)-\hbar\omega_L/2.
\label{ch5:eq:dressed-pot}
\end{eqnarray}
\noindent
Expanding, at each time $t$, the wavepackets $\Psi_g(R,t)$ and $\Psi_e(R,t)$ on the stationary vibrational levels $\varphi_{g,v"}$ and  $\varphi_{e,v'}$ of the $g$ and $e$ electronic states (Eq.~(\ref{ch5:eq:ampli-wp})), we obtain the instantaneous amplitude of population, $a_{v"}(t)$ (resp. $b_{v'}(t)$), in the stationary levels $g \,v"$ (resp. $e \, v'$), with energies $E_{g,v"}$  (resp. $E_{e,v'}$):
\begin{eqnarray}
a_{v"}(t)\exp\left[-i\frac{E_{g,v"}}{\hbar} t\right]&=&\int_0^L \varphi_{g,v"}(R) \Psi_g(R,t) dR , \nonumber \\
b_{v'}(t)\exp\left[-i\frac{E_{e,v'}}{\hbar} t\right]&=&\int_0^L \varphi_{e,v'}(R) \Psi_e(R,t) dR.  \nonumber \\
\label{ch5:eq:a,b-wp}
\end{eqnarray}
\subsection{\label{app:LbyL} Laser-coupled vibrational levels: 'Level by Level' description}
\noindent
In the LbyL description, the time-dependent wave function $\underline{\Psi}(t)$ is decomposed on the sets ${\underline{g}}_n$ (resp. ${\underline{e}}_m$), with $g_n$ (resp. $e_m$) wave functions $\varphi_{g,v"}$ (resp.  $\varphi_{e,v'}$) describing  stationary vibrational levels of the ground and excited electronic states. In the interaction picture \cite{bookCohen}, the expression of the wavepackets created by the laser pulse on the ground and excited states are given by  Eq.~(\ref{ch5:eq:ampli-LbyL}).

\noindent
The time-dependent Schr\"{o}dinger equation governing the time evolution of the ground $\underline{a}_{v"}(t)$ and excited $\underline{b}_{v'}(t)$  probability amplitudes is equivalent to the system of $(g_n+e_m)$ coupled equations:
\begin{eqnarray}
\stackrel{.}{\underline{a}}_{v"}& = &-\frac{i}{\hbar} \sum_{v'\in \underline{e}_m} {\underline{b}}_{v'} \exp{\left[-\frac{i}{\hbar}(E_{e,v'}-E_{g,v"})t\right]} \nonumber \\
 &\times& \ \ \ \ \  \left\langle\varphi_{g,v"} \left|\mbox{${\hat W}$}_{ge}\right|\varphi_{e,v'}\right\rangle , \nonumber \\
\stackrel{.}{\underline{b}}_{v'}& = &-\frac{i}{\hbar} \sum_{v"\in \underline{g}_n} {\underline{a}}_{v"} \exp{\left[-\frac{i}{\hbar}(E_{g,v"}-E_{e,v'})t\right]} \nonumber \\
&\times& \ \ \ \ \ \left\langle\varphi_{e,v'} \left|\mbox{$\hat{W}$}_{eg}\right|\varphi_{g,v"}\right\rangle . 
\label{ch5:eq:eq-coupl}
\end{eqnarray}

\noindent
In this system, there appear only the off-diagonal matrix elements of the coupling $\mbox{$\hat{W}_{ge}(t)$}=-D_{ge}{\cal{E}}_0f(t)\cos[\omega_Lt]$. 
In the RWA approximation, when the high frequency   $[(E_{e,v'}-E_{g,v"})/\hbar+\omega_L]\sim 2\omega_L$ and $[(E_{g,v"}-E_{e,v'})]/\hbar-\omega_L\sim -2\omega_L$ can be neglected, the system reduces to:
\begin{eqnarray}
\stackrel{.}{\underline{a}}_{v"}& = & - \frac{i}{2\hbar}  \sum_{v'\in\underline{e}_m} {\underline{b}}_{v'} \exp{\left[\frac{i}{\hbar} \delta_{v',v"} t\right]} \left\langle \varphi_{g,v"}|{W}_{ge}|\varphi_{e,v'}\right\rangle , \nonumber \\
\stackrel{.}{\underline{b}}_{v'}& = & - \frac{i}{2\hbar}  \sum_{v"\in\underline{g}_n} {\underline{a}}_{v"} \exp{\left[-\frac{i}{\hbar} \delta_{v',v"} t\right]} \left\langle \varphi_{e,v'}|{W}_{eg}|\varphi_{g,v"}\right\rangle . \nonumber \\
\label{ch5:eq:eq-coupl-RWA}
\end{eqnarray}

\noindent
 $W_{ge}(R,t)=W_{eg}(R,t)$ is defined in Eq.~(\ref{ch5:eq:W}) and $\delta_{v',v"}$ is equal to:
\begin{equation}
\delta_{v',v"}=\hbar\omega_L - [E_{e,v'}-E_{g,v"}] .
\label{ch5:eq:delta}
\end{equation}

\noindent
If the $R$-variation of  the electric dipole moment
${D}_{ge}({R})$ is neglected, introducing   $\mathbf{D}={D}_{ge}({R\rightarrow\infty} )$, one has:
\begin{eqnarray} 
 \left\langle e v'|W_{eg}(R,t)|g v" \right\rangle &=& \left\langle g v" |W_{ge}(t)|e v'\right\rangle \nonumber \\
 = -\mathbf{D} {\cal{E}}_0 f(t) \left\langle v'|v"\right\rangle 
 &=&\Omega_{v',v"} f(t).
 \label{ch5:eq:rabi}
\end{eqnarray}
 
\noindent
We solved the differential equations (\ref{ch5:eq:eq-coupl-RWA}) using the function {\it NDSolve} of the Mathematica software system.  
\noindent
The energies and wave functions for the levels $v"$ and $v'$, as well as the overlap integrals $\left\langle v'|v"\right\rangle$, were obtained from the MFGH  method (see Appendix \ref{app:MFGH}). 
\subsection{\label{app:WP-LbyL} 'Wavepacket' and 'Level by Level' descriptions} 

\noindent
Using the expansion Eq.~(\ref{ch5:eq:ampli-wp}) of the wavepackets $\chi_g(R,t)$ and $\chi_e(R,t)$ in terms of the stationary wave functions, and accounting for the closure relations satisfied by the  wave functions $\varphi_{g,v"}$ and $\varphi_{e,v'}$ (Eq.~(\ref{ch5:eq:ortho})), one obtains a system of $2{N}$  first-order differential equations for the probability amplitudes of  vibrational states $a_{v"}(t)$ and $b_{v'}(t)$ involved in the WP description. This system  is very similar to the system Eq.~(\ref{ch5:eq:eq-coupl}) satisfied by the probability amplitudes  ${\underline{a}}_{{v"}}(t)$ and ${\underline{b}}_{{v'}}(t)$ in the LbyL description.

\noindent
The difference between the two systems arises only from the number of involved amplitudes : $2{N}$ for the WP description and $(g_n+e_m)$ in the  LbyL approach. We emphasize that the WP description takes automatically advantage of the completeness character of the set of eigenfunctions provided by the spatial representation of the Hamiltonian on a grid.  The description of the dynamics does not depend on the choice of the grid parameters, provided that a sufficiently wide energy range is spanned by the eigenvalues obtained in the diagonalization. Thus the WP method provides a general non-perturbative treatment of the molecule-laser interaction, limited to the considered electronic states. It is straightforward to extend the two-states model employed  here to models with several electronic states. Such multi-surface models may become necessary, for example, in studies of  photoexcitation of vibrational levels belonging to electronic states coupled by molecular interactions.   
\section{\label{app:multilevel}RWA, diabatic and adiabatic basis, adiabaticity}
\subsection{\label{app:subsec:LbyL:RWA-laser-wt} RWA at the laser frequency, diabatic basis} 

\noindent
The interaction picture has been used, in the LbyL framework (Sec. \ref{app:LbyL}), to analyze the dynamics of the vibrational population transfer. In this approach, the Hamiltonian is non-diagonal, with matrix elements including  terms $\exp[\pm i \delta_{v',v"} t ]$ (Eq.~(\ref{ch5:eq:eq-coupl-RWA})), with oscillating contributions depending on the detuning of the laser with respect to the frequency of the $gv"\rightarrow ev'$ transition.

\noindent
Instead of working in the interaction picture, one may transform into a reference frame rotating at the laser frequency $\omega_L/2\pi$. 
The laser is resonant with the transition $g\,v_0" \rightarrow e\,v_0'$. The time-dependent wave function $\Psi(t)$ is explicitly expanded over the {\em diabatic} basis made of $g_n$ wave functions $|{\cal{E}}_g\,v"\rangle$ of vibrational levels in the ground electronic state and $e_m$ wave functions $|{\cal{E}}_e\,v'\rangle$ of levels of the excited electronic state:
\begin{eqnarray}
\Psi(t)&=&\exp(-\frac{i}{\hbar}E_{g,v_0"}t) \sum_{v"\in\underline{g}_n} \underline{A}_{v"}(t) \, |{\cal{E}}_g\,v"\rangle  \\
&+&
\exp(-\frac{i}{\hbar}E_{g,v_0"}t) \sum_{v'\in\underline{e}_m } \underline{B}_{v'}(t) \,\exp(-i \omega_L t)|{\cal{E}}_e\,v'\rangle. \nonumber 
\label{ch5:eq:rot-frame}
\end{eqnarray}
\noindent
In the RWA approximation, {\it{i.e.}} neglecting the rapidly oscillating terms $\exp(-i2\omega_L t)$ (resp. $\exp(+i2 \omega_L t)$), the amplitudes  ${\underline{A}}_{v"}$ and ${\underline{B}}_{v'}$ satisfy the system of coupled first-order differential equations:
{\small
\begin{eqnarray}
i \stackrel{.}{\underline{A}}_{v"}& = &  \Delta^g_{v",v_0"} \; {\underline{A}}_{v"} + \frac{1}{2} \sum_{v'\in\underline{e}_m} \Omega_{v',v"} f(t){\underline{B}}_{v'} , \nonumber \\
i \stackrel{.}{\underline{B}}_{v'}& = &   \Delta^e_{v',v_0'}{\underline{B}}_{v'} + \frac{1}{2} \sum_{v"\in\underline{g}_n}  \Omega_{v',v"} f(t) \; {\underline{A}}_{v"} ,
\label{ch5:eq-coupl-rot-frame}
\end{eqnarray}
}
\noindent
where $\Omega_{v',v"}$ are given by Eq.~(\ref{ch5:eq:rabi}) and where the energy differences $\Delta^{g \,\mathrm{or} \,e}_{v_1,v_2}$ are defined in Eq.~(\ref{ch5:eq:Delta-g-e}).

\noindent
The time-dependent Hamiltonian $\hat{{\cal{H}}}_{diab}(t)$ is represented in the diabatic basis by the following matrix:
{\footnotesize 
\begin{eqnarray}
\hat{{\cal{H}}}_{diab}(t)= \hspace{5cm} \nonumber \\
\left(
\overbrace{\begin{array}{lcc}
 \Delta^g_{v",v_0"}&0&0 \\ \\
0&\Delta^g_{v",v_0"}&0 \\ \\
0&0&\Delta^g_{v",v_0"}  \\
\ddots&\ddots&\ddots\\
\ddots&\frac{1}{2}\Omega_{v',v"} f(t)&\ddots \\
\ddots&\ddots&\ddots
\end{array}}^{g_n}
\overbrace{\begin{array}{lcc}
\ddots&\ddots&\ddots\\
\ddots&\frac{1}{2}\Omega_{v',v"} f(t)&\ddots \\
\ddots&\ddots&\ddots \\
\Delta^e_{v',v_0'}&0&0\\ \\
0&\Delta^e_{v',v_0'}&0\\ \\
0&0&\Delta^e_{v',v_0'}
\end{array}}^{e_m} \right)
 \label{ch5:eq:h(t)}
\end{eqnarray}
} 
\subsection{\label{app:subsec:adiab}Instantaneous adiabatic basis} 

\noindent
At each time $t$, the diabatic time-dependent Hamiltonian  $\hat{{\cal{H}}}_{diab}(t)$ can be diagonalized, determining the $(g_n+e_m)$ field-dressed or {\em adiabatic} levels $\left|j(t)\right\rangle$, with eigenvalues ${\epsilon}_j(t)$  \cite{bookMessiah}:
\begin{equation}
\hat{{\cal{H}}}_{diab}(t) \left|j(t)\right\rangle =  \epsilon_j(t)\left|j(t)\right\rangle .
\label{ch5:eq:adiab}
\end{equation}

\noindent
These states can be considered as a family of solutions of the time-independent Schr\"{o}dinger equation, with the time $t$ as a parameter. The  normalization condition is $\left\langle j(t)|j(t)\right\rangle=1$ and the integral $\left\langle j(t)|\stackrel{.}{j(t)}\right\rangle$, is thus purely imaginary. The phase of each eigenvector can be chosen arbitrarily at each time $t$, and it is possible to choose the phase in such a way that $\left\langle j(t)|\stackrel{.}{j(t)}\right\rangle=0$ \cite{bookSchiff}.

\noindent
The  diabatic Hamiltonian  can equivalently be written in the adiabatic basis ${\hat{{\cal{H}}}}_{adiab}(t)\equiv {\hat{{\cal{H}}}}_{diab}(t)$, with:
\begin{equation}
\hat{{\cal{H}}}_{adiab}(t)= \sum_{j=1}^{g_n+e_m} \epsilon_j(t) \left|j(t)\right\rangle \left\langle j(t)\right| ,
\label{ch5:eq:hamil-tot}
\end{equation}

\noindent
and the wave function $\Psi(t)$ can be decomposed on the adiabatic basis:
\begin{equation}
\Psi_{}(t)= \sum_{j=1}^{g_n+e_m} e_j(t) \left|j(t)\right\rangle .
\label{ch5:eq:adiab-dec}
\end{equation}

\noindent
The amplitudes of population $e_j(t)$ of the instantaneous adiabatic levels $\left|j(t)\right\rangle$ obey the following system of $(g_n+e_m)$ coupled equations:
\begin{equation}
i\hbar \frac{d \, e_j(t)}{dt}={ \epsilon}_j(t) e_j(t)- i\hbar \sum_{k=1}^{g_n+e_m} \alpha_{j,k}(t) e_k(t) .
\label{ch5:eq:adiab-evol}
\end{equation}

\noindent
The coefficient $\alpha_{j,k}(t) =\left\langle j(t) | \stackrel{.}{k(t)} \right\rangle=-\left\langle \stackrel{.}{j(t)}|{k}(t)\right\rangle$ describes the variation of the adiabatic level $\left|k(t)\right\rangle$ in the adiabatic basis \cite{bookMessiah}. With the particular phase convention written above \cite{bookSchiff},  the sum over $k$ in Eq.~(\ref{ch5:eq:adiab-evol}) does not include $k=j$.

\noindent
An expression of $\alpha_{j,k}(t)$ for $k\neq j$ is
\begin{equation}
\left[\epsilon_k(t) -\epsilon_j(t)\right] \alpha_{j,k}(t) = \left\langle j(t)\right|\frac{\partial{\hat{\cal{H}}}_{adiab}}{\partial t}\left|k(t)\right\rangle - \frac{\partial\epsilon_k(t)}{\partial t} \delta_{kj} .
\label{ch5:eq:angular}
\end{equation}
\subsection{{\label{app:subsubsec:adiab-approx}}Adiabatic approximation}

\noindent
In the adiabatic approximation, the second term on the r.h.s. of Eq.~(\ref{ch5:eq:adiab-evol}) is neglected, and the adiabatic amplitudes evolve as:
\begin{eqnarray} 
e_j(t)=e_j(t=0) \exp\left[-\frac{i}{\hbar} \int_0^{t} \epsilon_j(t') dt'\right]. 
\label{ch5:eq:adiab-state}
\end{eqnarray}
In this approximation, when the system is at the initial time in an instantaneous eigenstate of the Hamiltonian at $t=0$, let us say $\left|j_0(t=0)\right\rangle$, {\it{i.e.}} when in Eq.~(\ref{ch5:eq:adiab-dec}) $e_j(t=0)=\delta_{j,j_0}$, the system remains in the instantaneous eigenstate that evolves from the initial one, and there is no jump toward different instantaneous adiabatic states.

\noindent
The validity of the adiabatic approximation has been discussed in several papers \cite{bookMessiah,mac-kenzie2006,mac-kenzie2007,tong2007,wei2007}. From Messiah \cite{bookMessiah}, a condition of validity is given by:
\begin{equation}
\frac{\max[\sum_k  \left\langle \stackrel{.}{j}_k(t)|j_0(t)\right\rangle^2]}{\min[|(\epsilon_k(t)-\epsilon_{j_0}(t))/\hbar|^2]}\ll 1,
\end{equation}
\noindent
but this condition is clearly questionable \cite{bookTannor}
and other criteria are given, such as
\begin{equation}
\left|\frac{\left\langle \stackrel{.}{j}_k(t)|j_m(t)\right\rangle}{\epsilon_j(t)-\epsilon_m(t)}\right|\ll1, \; \; \rm{or} \; \left|\frac{\left\langle {j}_k(t)|\frac{\partial \cal{H}}{\partial t}|j_m(t)\right\rangle}{[\epsilon_j(t)-\epsilon_m(t)]^2}\right|\ll1.
\end{equation}

\noindent  
In fact, there is no fully general rule allowing to predict the validity of the adiabatic approximation \cite{bookMessiah}. The more or less adiabatic character of the evolution can be discussed {\em a posteriori}, once the evolution of the system has been computed, by studying the evolution of the population in different adiabatic levels. 
\subsection{\label{app:subsec:diab-2-level} Case of a two-level system}

\noindent
In the diabatic basis describing the two-level system $[\Psi_g,\;\Psi_e]$, the effective time-dependent Hamiltonian in the RWA approximation is (Eq.~(\ref{ch5:eq:h(t)})): 
\begin{equation}
H_{diab}(t) = \left[\begin{array}{lc}
 +0& 
 \Omega(t) \\ 
 \Omega(t) & 
 \delta
 \end{array} \right] ,
\label{ch5:eq:definition-Diab}
\end{equation}
where $\delta$ denotes the detuning of the laser excitation, and $\Omega(t)=\frac{1}{2}\Omega_{e,g}f(t)$ is the instantaneous coupling. The probability amplitudes in the diabatic basis satisfy the first-order differential system :  
\begin{eqnarray}
i\dot{a}(t) & = & \Omega(t) b(t) , \nonumber \\
i\dot{b}(t) & = & \delta b(t) + \Omega(t) a(t) .
\label{ch5:eq:2-level-diabatic}
\end{eqnarray}

\noindent
The population transferred to the excited level at time $t$ reads
\begin{equation}
P_{diab}(t)=|b(t)|^2 .
 \label{ch5:eq:diabaticB-population}
\end{equation}

\noindent
The adiabatic character of the process can be analyzed by introducing the instantaneous adiabatic basis. This basis, $[\Psi_-,\;\Psi_+]$, can be obtained from the unitary matrix ${\hat{U}}(t)$ that diagonalizes $H_{diab}(t)$ defined in Eq.~(\ref{ch5:eq:definition-Diab}) \cite{bookCohen}:
\begin{equation}
{\hat{U}}(t)= \left[\begin{array}{lc}
 \cos \frac{\theta}{2} e^{-i\phi/2}& 
 -\sin \frac{\theta}{2} e^{-i\phi/2} \\
 \sin \frac{\theta}{2} e^{i\phi/2} & 
 \cos \frac{\theta}{2} e^{i\phi/2}
 \end{array} \right], 
\label{ch5:eq:unitaryMatrix}
\end{equation}

\noindent
with eigenvalues
\begin{equation}
 E_{\pm}(t)=\frac{1}{2}\delta \pm \frac{1}{2}\sqrt{ \delta^2 +4 \left|\Omega(t)\right| ^2 } .
 \label{ch5:eq:adiabatic-e}
\end{equation}

\noindent
$\theta(t)$ is defined by the relation
\begin{equation}
\tan \theta(t) = \frac{2 \left| \Omega(t) \right|}{\delta}, \ \ 0 \leq \theta < \pi ,
\label{ch5:eq:angle-rotation}
\end{equation}

\noindent
and the phase $\phi(t)$ by
\begin{equation}
\Omega(t)=\left|\Omega(t)\right|e^{i\phi(t)}, \ \ 0\leq \phi < 2\pi .
 \label{ch5:eq:complex-coupling}
\end{equation}

\noindent
For the excitation by an unchirped Gaussian pulse, one can take $\phi(t)\equiv0$. The instantaneous adiabatic states $\Psi_-(t)$ and $\Psi_+(t)$ are related to the diabatic ones by:
\begin{equation}
 \left(\begin{array}{c}
  \Psi_-(t)  \\
  \Psi_+(t)
  \end{array} \right)=\hat{U}^{-1}(t)\left(\begin{array}{c}
  \Psi_g \\
  \Psi_e
  \end{array} \right) .
\label{ch5:eq:adiabatic-diabatic}
\end{equation}

\noindent
They are solution of 
\begin{eqnarray}
i&\hbar&\frac{d}{dt}\left(\begin{array}{c}
 \Psi_- \\
 \Psi_+
 \end{array} \right)  =   \left(\begin{array}{lc}
 E_- & 0 \\
 0 & E_+
 \end{array} \right) \left(\begin{array}{c}
 \Psi_-   \\
 \Psi_+
 \end{array} \right) \\
 & + & \left[ -\frac{i\hbar\dot{\theta}}{2}\left(\begin{array}{lc}
 0 & -1  \\
 1 & 0
 \end{array} \right) -\frac{\hbar \dot{\phi}}{2}\left(\begin{array}{lc}
 \cos\theta & -\sin\theta \\%
 -\sin\theta & -\cos\theta
 \end{array} \right) \right] \left(\begin{array}{c}
 \Psi_-  \nonumber \\
 \Psi_+
 \end{array} \right) .
\label{ch5:eq:schrodinger-adiabatic-d}
\end{eqnarray}

\noindent
In the adiabatic basis, the solution of the time-dependent Schr\"odinger equation can be written as 
\begin{equation}
 \Psi_{}(t)=e^{-i\frac{\delta}{2}t}\left[\alpha(t)\left|\Psi_-(t)\right\rangle + \beta(t) \left|\Psi_+(t)\right\rangle\right] \, . 
 \label{ch5:eq:Psi-adia}
 \end{equation}
 
\noindent
The coupled system for the amplitudes of the instantaneous adiabatic levels is: 
\begin{eqnarray}
i\dot{\alpha} & = & -\frac{1}{2}\sqrt{\delta^2+4 \Omega^2}\;\alpha -\frac{i}{2}\stackrel{\cdot}{\theta}(t)\beta , \nonumber \\
i\dot{\beta} & = & \frac{1}{2}\sqrt{\delta^2+4 \Omega^2}\;\beta + \frac{i}{2}\stackrel{\cdot}{\theta}(t)\alpha .
\label{ch5:eq:2-level-adiabatic}
\end{eqnarray}

\noindent
The population in the instantaneous adiabatic levels $\left|\Psi_-(t)\right\rangle$ and $\left|\Psi_+(t)\right\rangle$ may be found as
\begin{eqnarray}
P_{adiab}^{(+)}(t)=|\beta(t)|^2 , \nonumber \\
P_{adiab}^{(-)}(t)=|\alpha(t)|^2 .
 \label{ch5:eq:adiabaticB-population}
\end{eqnarray}

\noindent
The second term on the right hand side of Eq.~(\ref{ch5:eq:2-level-adiabatic}) represents the non-adiabatic coupling between the adiabatic levels. 
When in Eq.~(\ref{ch5:eq:unitaryMatrix}) $\phi(t)\equiv0$, the non-adiabatic coupling is proportional to $\frac{\hbar\dot{\theta}}{2}$. The non-adiabatic coupling can be neglected if and only if
\begin{equation}
 \frac{1}{4}\left(\hbar \dot{\theta} \right)^2 
 \ll \left(E_+ - E_-\right)^2 ,
 \label{ch5:eq:condition-nonadiabatic}
\end{equation}

\noindent
The evolution is then adiabatic and the instantaneous adiabatic levels evolve as:
\begin{equation}
\Psi_\pm (t) = \Psi_\pm (t=0) \exp{(-\frac{i}{\hbar} \int_0^t E\pm(t') dt' )}
\label{ch5:eq:adiabaticity-wf}
\end{equation}
At $t=0$, we assume that $\alpha(t=0)=1$ and $\beta(t=0)=0$ or, equivalently, that $\Psi(t=0)=\Psi_-(t=0)=\Psi_g$. In the adiabatic approximation, the probability amplitudes  in the diabatic basis, the so-called 'Rabi oscillations' \cite{bookCohen} can be simply calculated by using the general prescription of Sec.~IV~C~3a of Ref.~\cite{bookCohen}: 

1) at $t=0$,  project the initial probabilty amplitude, defined in the diabatic basis, onto the adiabatic basis,  using the transformation 
(Eq.~(\ref{ch5:eq:adiabatic-diabatic})).

2) 
propagate  adiabatic states according to Eq.~(\ref{ch5:eq:adiabaticity-wf}).

3) project $\Psi(t)$ on the diabatic basis.
\section{\label{ch5:subsec:N-levels} Resonant excitation from a $N$-fold degenerate level}

\noindent 
We consider the excitation from a system of N degenerate levels $\left|g\,v"\right\rangle$, $v"=1$ to $N$, toward a single level $\left|e\,v_0'\right\rangle$ ($g_n=N\; e_m=1$).
The relevant Hamiltonian ${\cal{H}}_{diab}(t)$ (Eq.~(\ref{ch5:eq:h(t)}))  can be written as:
\begin{eqnarray}
{\cal{H}}_{diab}(t)= \hspace{5cm} \nonumber \\
\left(
\overbrace{\begin{array}{lccc}
0&0&\cdots&0 \\ 
0&0&\cdots&0 \\
\vdots&\vdots&\vdots&\vdots\\
0&0&\cdots&0\\
W(t)/2&W(t)/2&\cdots&W(t)/2
\end{array}}^{N}
\overbrace{
\begin{array}{c}
W(t)/2 \\
W(t)/2 \\ 
\vdots \\
W(t)/2 \\
\Delta 
\end{array}}^{1} \right) ,
\label{ch5:eq:h-N(t)}
\end{eqnarray}

\noindent
where $\Delta$ is a common detuning  and we assume that all couplings are equal, $W(t)=\Omega_{1,v"}f(t)$ for $v"=1$ to $N$.

\noindent
The energies of the adiabatic {levels} are given by:
\begin{eqnarray}
\underline{\epsilon}_j(t)&=&0 \;\rm{for} \; 1 \le j \le(N-1), \\
\underline{\epsilon}_{-}(t)=\underline{\epsilon}_{j=N}(t)&=&\frac{1}{2}\left[\Delta-\sqrt{\Delta^2 +N W^2(t)}\right],\nonumber \\
\underline{\epsilon}_{+}(t)=\underline{\epsilon}_{j=N+1}(t)&=&\frac{1}{2}\left[\Delta+\sqrt{\Delta^2 +N W^2(t)}\right] . \nonumber
 \label{ch5:eq:adiaEner-deg}
\end{eqnarray}

\noindent
The instantaneous adiabatic levels $|j(t)\rangle$ are defined by their components $V_{i,j}(t)$ on the diabatic levels $|i\rangle$, which satisfy
\begin{eqnarray}
\underline{\epsilon}_j(t) V_{i,j}&=& \frac{W(t)}{2} V_{N+1,j} \;\rm{for} \; 1 \le i \le N \nonumber \\ 
\frac{W(t)}{2}[V_{1,j}+ ..... + V_{N,j}] &=& [\underline{\epsilon}_j(t)- \Delta]\, V_{N+1,j}
\label{ch5:eq:V}
\end{eqnarray}

\noindent
For the degenerate eigenvalues $\underline{\epsilon}_j(t)$, with $1\le j\le (N-1)$, orthogonal eigenvectors can be found.  For example, we may construct the following orthogonal basis
{\small
\begin{eqnarray}
V_{i,j}(t)&=&-\frac{1}{\sqrt{j(j+1)}}  \; \; {\mathrm{for}} \; \; 1 \le i\le j \; \; {\mathrm{and}} \;\; 1\le j \le (N-1) \nonumber \\
V_{j+1,j}(t)&=&\sqrt{\frac{j}{j+1}}  \\
V_{i,j}(t)&=& 0 \; \; {\mathrm{for}} \; \; (j+2) \le i\le (N+1) .\nonumber
\label{ch5:eq:V-value1}
\end{eqnarray}
}

\noindent
For the two other eigenvalues $\underline{\epsilon}_\mp(t)$, the eigenvectors are determined by:

\noindent 
for $j=N$ and $j=N+1\;$: 
$\;\;V_{i,j}=V_{g,\mp}\;$, for $\;1 \le i \le N$  

\noindent
and $\;\;V_{(N+1),j}=V_{e,\mp}\; $, 
with $\;\;\underline{\epsilon}_\mp \,V_{g,\mp}=\frac{W}{2}\,V_{e,\mp}$ 

\noindent
and the normalization condition:
 
\noindent
$\;\;V_{g,\mp}^2 [N + (\frac{-\Delta \mp\sqrt{\Delta^2+N W^2(t)}}{W(t)})^2]=1$.  

\noindent
For resonant excitation ($\Delta=0$) one has $V_{g,\mp}=\frac{1}{\sqrt{2N}}$ and $V_{e,\mp}=\mp\frac{1}{\sqrt{2}}$.

\noindent 
We consider now the resonant excitation from a particular sublevel of the $N$-fold degenerate level of the ground state.
If the conditions of adiabatic evolution are satisfied, 
the evolution of the amplitudes of population $\underline{A}_i(t)$ and $\underline{B}_1(t)$ in the diabatic levels $\left|g,i\right\rangle$ and $\left|e,1\right\rangle$ can be calculated by using the general method described in Appendix \ref{app:subsec:diab-2-level}, because the instantaneous eigenvectors of the Hamiltonian in Eq.~(\ref{ch5:eq:h-N(t)}) are known. 
At $t=0$, the initial wave function $\left|g,1\right\rangle$ is expanded over the adiabatic levels $|j(t=0)\rangle$ (Eq.~(\ref{ch5:eq:adiab-dec})) leading to the initial amplitudes  $V_{i=1,j}(t=0)$. Each adiabatic level evolves then according to Eq.~(\ref{ch5:eq:adiab-state}). The amplitude of population of the diabatic level $|i\rangle$ at time $t$ is equal to
$$<i|\Psi(t)\rangle=\sum_j V_{i=1,j}(0) V_{i,j}(t) \exp[\int_0^t-\frac{i}{\hbar}\underline{\epsilon}_j(t')dt']\; .$$

\noindent
{Using Eqs.~(\ref{ch5:eq:V-value1}), one obtains the probability amplitudes in the ground state $\underline{A}_i(t)$ or in the excited state $\underline{B}_1(t)$:}
\begin{eqnarray}
  \underline{A}_1(t) &=& \frac{N-1}{N} + \frac{1}{N} \cos\left[\frac{\sqrt{N}}{2}\Theta(t)\right], \nonumber \\
  \underline{A}_i(t) &=& - \frac{1}{N} + \frac{1}{N} \cos\left[\frac{\sqrt{N}}{2}\Theta(t)\right] \; \; {\rm for} \; 2\le i \le N ,\nonumber \\
  \underline{B}_1(t) &=& \frac{1}{\sqrt{N}} {\sin}\left[\frac{\sqrt{N}}{2}\Theta(t)\right] , \nonumber \\
  {\rm where} \; \; \Theta(t)&=& \int_0^t W(t') dt' .
\label{ch5:eq:a-Deg1}  
\end{eqnarray}
  
\noindent  
Notice that time-evolution of a resonantly-excited two-level system is fully characterized by $\Theta(t)$ (Eqs.~(\ref{ch5:eq:rabi-res-2levels},\ref{ch5:eq:a-Deg1})).  By contrast, the evolution of  our $(N+1)$-level system is governed by the angle $\sqrt{N} \Theta(t)$. 
The absolute value of the probability amplitude of the excited level is reduced by the factor $\sqrt{N}$ and the population is redistributed among N levels of the ground state.   For a large number of degenerate levels ($N \rightarrow \infty$) and in the high field regime ($\Theta(t)$ large), the excitation process is blocked and the population remains in the initial level.

\noindent  
Conversely, in the low-field limit $\Theta(t) \ll \pi$, the population in the excited level in the $(N+1)$-level system, $\underline{B}_1(t)\sim \frac{1}{2} \Theta(t)$, is  weak and identical to the amplitude of population in a two-level system resonantly excited by the same pulse. Simultaneously, in the ground electronic state, there is no change in the amplitude of population of the levels $  \underline{A}_1(t)\sim 1$ and $   \underline{A}_i(t)\sim 0$ for $ 2\le i \le N$.
\section{\label{app:pulse-train} Ultrashort pulse train}

\noindent
The time-dependent  electric field describing a coherent train of Gaussian pulses is given by \cite{cundiff2002} 
\begin{eqnarray}
{\cal E}(t) &=& \frac{1}{2}\sum_{q=0}^{\mathcal{N}-1}{{\cal E}_0\exp{\left(i\Phi_0\right)}f\left(t-qT_{rep}\right)}  \nonumber \\
&\times& {\exp\left[i\left(\omega_L(t-qT_{rep})+q\Delta\phi_{ce}\right)\right]} ,
\label{ch5:eq:train-P}
\end{eqnarray}
where $\mathcal{N}$ is the number of pulses, $\Phi_0$ is a constant phase, $f$ is a Gaussian envelope given by Eq.~(\ref{ch5:eq:envelop}), $T_{rep}$ is the pulse repetition time, and $\Delta\phi_{ce}$ is the pulse-to-pulse carrier-envelope offset phase shift.

\noindent
In the energy domain, the spectral distribution of ${\cal E}(t)$ is obtained from the Fourier transform of Eq.~(\ref{ch5:eq:train-P}) 
\begin{eqnarray}
\tilde{{\cal E}}(\omega) & = & \frac{{\cal E}_0}{2}\tilde{f}\left(\omega-\omega_L\right)\exp\left[i\frac{\mathcal{N}-1}{2}\left(\Delta\phi_{ce}-\omega T_{rep}\right)\right] \nonumber \\
& \times & \exp{\left(i\Phi_0\right)}\frac{\sin{\left[\mathcal{N}\left(\Delta\phi_{ce}-\omega T_{rep}\right)/2\right]}}{\sin{\left[\left(\Delta\phi_{ce}-\omega T_{rep}\right)/2\right]}}.
\label{ch5:eq:train-P-freq}
\end{eqnarray}

\noindent
This distribution consists of a comb of  structures (``teeth'') located at equally-spaced frequencies  
$$f_m=mf_{rep}+\delta,$$
with spacing $f_{rep}=1/T_{rep}$; $\delta$ is the offset frequency equal to $\delta=\Delta\phi_{ce} f_{rep}/(2\pi)$. Each peak $m=0,1,2$ etc, has a maximum of  intensity  $[\tilde{f}(\omega_m-\omega_L)]^2 \mathcal{N}^2$ increasing as the square of the number of pulses. Here $\omega_m=2\pi f_m$ and  $\tilde{f}(\omega)$ denotes the Fourier Transform of $f(t)$. The width of each tooth $f_{rep}2\pi/\mathcal{N}$ decreases with increasing $\mathcal{N}$. 
\begin{acknowledgments}
The authors thank C. Koch, O. Dulieu and N. Bouloufa for stimulating discussions. This work was partially supported by grants from R\'egion Ile-de-France. The authors are grateful to the ECOS-NORD Program for funding the colaborative project C08P02 between Universidad de Antioquia and Laboratoire Aim\'e Cotton, Universit\'e Paris-Sud 11. B.L. thanks the office Relations Internationales de l'Universit\'e Paris-Sud 11 for financial support. Laboratoire Aim\'e Cotton, is unit\'e prope UPR 3321 of CNRS associe\'e \`a l'Universit\'e Paris-Sud 11, member of F\'ed\'eration Lumi\`ere Mati\`ere (LUMAT, FR 2764) and of the Institut Francilien de Recherche sur les Atomes Froids (IFRAF). The work of A.D. was supported in part by the U.S. Army Research Office and by the U.S. National Science Foundation.
\end{acknowledgments}
 

\begin{thebibliography}{}
\bibitem{kerman2004a} A. J. Kerman, J. M. Sage, S. Sainis, T. Bergeman, and D. DeMille, Phys. Rev. Lett. {\bf92}, 033004  (2004).
\bibitem{sage2005} J. M. Sage, S. Sainis, T. Bergeman, and D. DeMille, Phys. Rev. Lett. {\bf 94}, 203001 (2005).
\bibitem{bergeman2004} T. Bergeman, A. J. Kerman, J. M. Sage, S. Sainis, and D. DeMille, Eur. Phys. J. D. {\bf31}, 179 (2004).
\bibitem{stwalley2004} W. C. Stwalley, Eur. Phys. J. D {\bf31}, 221 (2004).
\bibitem{tscherneck2007} M. Tscherneck and N. P. Bigelow, Phys. Rev. A {\bf 75}, 055401 (2007).
\bibitem{ni2008} K.-K. Ni, S. Ospelkaus, M.H.G. de Miranda, A. Pe'er, B. Neyenhuis, J.J. Zirbel, S. Kotochigova, P.S. Julienne, D.S. Jin, and J. Je, Science {\bf 322}, 231 (2008).
\bibitem{luc2009} E. Luc-Koenig, and F. Masnou-Seeuws, in {\it Cold Molecules: Theory, Experiments and Applications}, edited by R. Krems, B. Friedrich, and W. Stwalley (CRC Press, Boca Raton, FL, 2009), p. 245.
\bibitem{luc2003} E. Luc-Koenig, R. Kosloff, F. Masnou-Seeuws, and M. Vatasescu, Phys. Rev. A {\bf 70}, 033414 (2004). 
\bibitem{luc2004} E. Luc-Koenig, F. Masnou-Seeuws, and M. Vatasescu, Eur. Phys. J. D {\bf 31}, 239 (2004). 
\bibitem{cundiff2002} S. T. Cundiff, J. Phys. D: Appl. Phys. {\bf 35} R43 (2002).
\bibitem{stowe2006} M. C. Stowe, F. C. Cruz, A. Marian, and J. Ye, Phys. Rev. Lett. {\bf 96}, 153001 (2006).
\bibitem{kokoouline1999} V. Kokoouline, O. Dulieu, R. Kosloff and F. Masnou-Seeuws, J. Chem. Phys. {\bf 110}, 9865 (1999).
\bibitem{willner2004} K. Willner, O. Dulieu and F. Masnou-Seeuws, J. Chem. Phys. {\bf 120}, 548 (2004).
\bibitem{londono2009} B. E. Londo\~{n}o, J. E. Mahecha, E. Luc-Koenig, and A. Crubellier, Phys. Rev. A {\bf 80}, 032511 (2009).
\bibitem{cao1998} J. Cao, Ch. J. Bardeen, and K. R. Wilson, Phys. Rev Lett. {\bf 80}, 1406 (1998).
\bibitem{cao2000} J. Cao, Ch. J. Bardeen, and K. Wilson, J. Chem. Phys {\bf 113}, 1898 (2000).
\bibitem{londono2011} B. E. Londo\~{n}o, J. E. Mahecha, E. Luc-Koenig, and A. Crubellier,   Phys. Chem. Chem. Phys. {\bf13}, 18738 (2011).
\bibitem{kosloff1994} R. Kosloff, Annu. Rev. Phys. Chem. {\bf 45}, 145 (1994).
\bibitem{bookCohen2} C. Cohen-Tannoudji, J. Dupont-Roc et G. Grynberg, {\it Processus d'interaction entre photons et atomes}, EDP Sciences (1996).
\bibitem{bookTannor} D. J. Tannor, {\it Introduction to quantum mechanics a time-dependent perspective}, University Science Books, California (2007).
\bibitem{shapiro2003} M. Shapiro, and P. Brumer, {\it Principles of the quantum control of molecular processes}, Wiley Interscience, New York (2003).
\bibitem{salzmann2008} W. Salzmann, T. Mullins, J. Eng, M. Albert, R. Wester, M. Weidem\"{u}ller, A. Merli, S. M. Weber, F. Sauer, M. Plewicki, F. Weise, L. W\"{o}ste, and A. Lindinger, Phys. Rev. Lett. {\bf 100}, 233003 (2008).
\bibitem{mccabe2009} D. J. McCabe, D. G. England, H. E. L. Martay, M. E. Friedman, J. Petrovic, E. Dimova, B. Chatel, and I. A. Walmsley, Phys. Rev. A {\bf 80}, 033404 (2009). 
\bibitem{bookMessiah} A. Messiah {\it M\'ecanique quantique Tome 2}, Dunod, Paris (1964).
\bibitem{albert2008} M. Albert, T. Mullins, S. G\"{o}tz, W. Salzman, R. Wester, and M. Weidem\"{u}ller, J. Mod. Opt. {\bf 55}, 3359 (2008).
\bibitem{merli2009} A. Merli, F. Eimer, F. Weise, A. Lindinger, W. Salzmann, T. Mullins, S. G\"{o}tz, R. Wester, M. Weidem\"{u}ller, R. A\~{g}ano\~{g}lu, and C. P. Koch, Phys. Rev. A {\bf 80}, 063417 (2009).
\bibitem{banin1994} U. Banin, A. Bartana, S. Ruhman, and R. Kosloff, J. Chem. Phys. {\bf 101}, 8461 (1994).
\bibitem{parker1990} J. Parker, and C. R. Stroud Jr., Phys. Rev. A {\bf41 }, 1602 (1990).
\bibitem{araujo1999} L. E. E. Araujo, and I. A. Walmsley, J. Phys. Chem. A {\bf 103}, 10409 (1999).
\bibitem{dubrovskii1992} Y. V. Dubrovskii, M. Y. Ivanov, and M. V. Fedorov, Laser Physics {\bf 2}, 288 (1992).
\bibitem{vitanov1995} N. V. Vitanov, and P. L. Knight, Phys. Rev. A {\bf 52}, 2245 (1995).
\bibitem{araujo2008} L. E. E. deAraujo, Phys. Rev. A {\bf 77}, 033419 (2008). 
\bibitem{thomas2010} S. Thomas, A. Malacarne, F. Fresi, L. Pot\`{i}, and J. Aza\~{n}a, J. of Lightwave Technology, {\bf 28} 1832 (2010).
\bibitem{weise2009} F. Weise, A. Merli, S. Birkner, F. Sauer, L. W\"{o}ste, A. Lindenger, R. Aganoglu, C.P. Koch, W. Salzmann, T. Mullins, S. G\"{o}tz, R. Wester, and M. Weidem\"{u}ller, Eur. Phys. J. D {\bf54}, 711 (2009).
\bibitem{shapiro2009} E. A. Shapiro, V. Milner, and M. Shapiro, Phys. Rev. A {\bf79}, 023422 (2009).
\bibitem{zhdanovich2009} S. Zhdanovich, E. A. Shapiro, J. W. Hepburn, M. Shapiro, and V. Milner, Phys. Rev. A {\bf 80}, 063405 (2009).
\bibitem{peer2007} A. Pe'er, E. A. Shapiro, M. C. Stowe, M. Shapiro, and J. Ye, Phys. Rev. Lett. {\bf98}, 113004 (2007).
\bibitem{bookCohen} C. Cohen-Tannoudji, B. Diu, and F. Lalo\"{e} {\it Mecanique Quantique Tome I}, Hermann, Paris (1977).
\bibitem{bookSchiff} L. I. Schiff {\it Quantum Mechanics}, McGraw-Hill, New York (1968).
\bibitem{mac-kenzie2006} R. MacKenzie, E. Marcotte, and H. Paquette, Phys. Rev. A {\bf 73}, 042104 (2006).
\bibitem{mac-kenzie2007} R. MacKenzie, A. Morin-Duchesne, H. Paquette, and J. Pinel, Phys. Rev. A {\bf 76}, 044102 (2007).
\bibitem{tong2007} D. M. Tong, K. Singh, L.C. Kwek, and C.H. Oh, Phys. Rev. Lett. {\bf98}, 150402 (2007).
\bibitem{wei2007} Z. Wei, and M. Ying, Phys. Rev. A {\bf 76}, 024304 (2007).
\end{thebibliography}
\end{document}